\documentclass[
 reprint,
preprint,
 amsmath,amssymb,
]{elsarticle}

\usepackage{amsthm}
\newtheorem{theorem}{Theorem}
\usepackage[export]{adjustbox}
\usepackage{xcolor}
\definecolor{C0}{HTML}{1F77B4}
\definecolor{C1}{HTML}{FF7F0E}
\definecolor{C2}{HTML}{2ca02c}
\definecolor{C3}{HTML}{d62728}
\definecolor{C4}{HTML}{9467bd}
\definecolor{C5}{HTML}{8c564b}

\def\usetodonotes{} 
\ifdefined\usetodonotes
\usepackage{todonotes}
\usepackage{changes}
\newcommand{\Ap}{\bm A^{\|}}
\newcommand{\Bp}{\bm B^{\|}}
\newcommand{\Cp}{\bm C^{\|}}
\newcommand{\Dp}{\bm D^{\|}}
\newcommand{\ap}{\bm a^{\|}}
\newcommand{\bp}{\bm b^{\|}}
\newcommand{\cp}{\bm c^{\|}}
\newcommand{\dpa}{\bm d^{\|}}

\fi

\usepackage{upgreek}
\usepackage{tikz}
\usepackage{soul}
\usepackage{tcolorbox}
\usepackage[scr=esstix,cal=boondox]{mathalfa}

\usepackage{graphicx}
\usepackage{dcolumn}
\usepackage{bm}
\usepackage[
margin=1in,
]{geometry}

\usepackage{siunitx}
\usepackage{siunitx}
\usepackage[skip=0pt]{caption}
\usepackage{subfig}
\usepackage{float}
\usepackage{amsmath}
\usepackage{amsmath}
\usepackage{amsfonts}
\newcommand{\angstrom}{\textup{\AA}}

\newcommand{\bcal}[1]{\bm{\mathcal #1}}

\usepackage{comment}
\usepackage[english]{babel}
\usepackage{natbib}
\setcitestyle{authoryear,comma,sort&compress}
\usepackage{algpseudocode}
\usepackage{algorithm}
\usepackage{latexsym}
\DeclareMathOperator{\divr}{Div}

\usepackage[colorlinks=true]{hyperref}
\hypersetup{ 
    colorlinks=true,       
    linkcolor=red,          
    citecolor=blue,        
    filecolor=magenta,      
    urlcolor=cyan           
}
\usepackage{cleveref}

\newcommand{\psubref}[1]{\protect\subref{#1}}
\newcommand{\ut}{\bm u_{\rm t}}
\newcommand{\ub}{\bm u_{\rm b}}
\newcommand{\up}{\bm u_+}
\newcommand{\um}{\bm u_-}

\newcommand{\Ft}{\bm F_{\rm t}}
\newcommand{\Fb}{\bm F_{\rm b}}
\newcommand{\Xt}{\bm X_{\rm t}}
\newcommand{\Xb}{\bm X_{\rm b}}

\newcommand{\xt}{\bm x_{\rm t}}
\newcommand{\xb}{\bm x_{\rm b}}
\newcommand{\Ht}{\bm H_{\rm t}}
\newcommand{\Hb}{\bm H_{\rm b}}
\newcommand{\Kt}{\bm K_{\rm t}}
\newcommand{\Kb}{\bm K_{\rm b}}

\newcommand{\phit}{\bm{\phi}_{\rm t}}
\newcommand{\phib}{\bm{\phi}_{\rm b}}
\newcommand{\rt}{\bm r_{\rm t}}
\newcommand{\rb}{\bm r_{\rm b}}

\newcommand{\thref}[1]{Theorem~\ref{#1}}
\newcommand{\sref}[1]{Section~\ref{#1}}
\newcommand{\aref}[1]{Algorithm~\ref{#1}}
\newcommand{\fref}[1]{Fig.~\ref{#1}}
\newcommand{\frefs}[2]{Figs.~\ref{#1} and \ref{#2}}
\newcommand{\divrt}{\divr}
\newcommand{\divrb}{\divr}

\begin{document}


\title{Bicrystallography-informed Frenkel--Kontorova model for interlayer dislocations in strained 2D heterostructures
}

\author[1]{Md Tusher Ahmed}
\author[2]{Chenhaoyue Wang}
\author[2]{Amartya S. Banerjee}
\author[1]{Nikhil Chandra Admal}%
\ead{admal@illinois.edu}
\affiliation[1]{%
 Department of Mechanical Science and Engineering
 University of Illinois at Urbana-Champaign, Champaign, IL, USA
}%
\affiliation[2]{%
Department of Materials Science and Engineering,
University of California, Los Angeles., Los Angeles, CA, USA
}%





\begin{abstract}
In recent years, van der Waals (vdW) heterostructures and homostructures, which consist of stacks of two-dimensional (2D) materials, have risen to prominence due to their association with exotic quantum phenomena originating from correlated electronic states harbored by them. Atomistic scale relaxation effects play an extremely important role in the electronic scale quantum physics of these systems, providing means of manipulation of these materials and allowing them to be tailored for emergent technologies. We investigate such structural relaxation effects in this work using atomistic and mesoscale models, within the context of twisted bilayer graphene --- a well-known heterostructure system that features moir\'{e} patterns arising from the lattices of the two graphene layers. For small twist angles, atomic relaxation effects in this system are associated with the natural emergence of interface dislocations or strain solitons, which result from the cyclic nature of the generalized stacking fault energy (GSFE), that measures the interface energy based on the relative movement of the two layers. In this work, we first demonstrate using atomistic simulations that atomic reconstruction in bilayer graphene under a large twist also results from interface dislocations, although the Burgers vectors of such dislocations are considerably smaller than those observed in small-twist systems. To reveal the translational invariance of the heterointerface responsible for the formation of such dislocations, we derive the translational symmetry of the GSFE of a 2D heterostructure using the notions of coincident site lattices (CSLs) and displacement shift complete lattices (DSCLs). The workhorse for this exercise is a recently developed Smith normal form bicrystallography framework. Next, we construct a bicrystallography-informed and frame-invariant Frenkel—Kontorova model, which can predict the formation of strain solitons in arbitrary 2D heterostructures, and apply it to study a heterostrained, large-twist bilayer graphene system. Our mesoscale model is found to produce results consistent with atomistic simulations. We anticipate that the model will be invaluable in predicting structural relaxation and for providing insights into various heterostructure systems, especially in cases where the fundamental unit cell is large and therefore, atomistic simulations are computationally expensive.
\end{abstract}

\maketitle


\section{\label{sec:level1}Introduction}
Quantum materials, i.e., materials that manifest exotic physical properties due to the presence of strong electronic correlations, have risen to prominence in recent years due to their applications in emergent technologies connected to nanoelectronics and quantum information science \citep{basov2017towards, keimer2017physics, tokura2017emergent}. The grand challenge of designing and manufacturing such materials stems from the high sensitivity of their properties to local structure and symmetry \citep{kim2022strain}. In recent years, van der Waals (vdW) homostructures and heterostructures, which consist of stacks of two-dimensional (2D) materials, have emerged as an important class of quantum materials \citep{Shimazaki_2020,Jin_2019,Cao2021,Regan_2020}. The weak vdW interactions between the 2D lattices in such materials offer high fidelity in tuning the local atomic environments, thus allowing exquisite control over the quantum properties of such systems. Small-twist bilayer graphene (BG) is the most prominent example, wherein dispersionless electronic states (or \textit{flat bands}) emerge \citep{Macdonald_2011, Origin_Magic_2019,cao_unconventional_2018,Tao_2022,Zhao_2020,Zhao_2021} at a specific \emph{magic} twist angle, $\theta \sim 1.1^\circ$. Magic-angle twisted BG exhibits unconventional superconductivity, correlated insulator phases, magnetism, and non-trivial topological phases \citep{PAPAGEORGIOU201775,lee_park_choi_2019,cao_2020,tawfiq_2022,uri2020,wong2020} --- properties associated with the \emph{moir\'e superlattice} formed by the constituent 2D lattices. However, such exotic properties are susceptible to perturbations in the twist angle. Since the relative twist between adjoining lattices constitutes only a one-dimensional subspace of the four-dimensional space of relative deformations,\footnote{If the two lattices of a heterostructure are subjected to uniform deformation gradients $\bm F_1$ and $\bm F_2$, then $\bm F_1^{-1}\bm F_2$ is the relative deformation, and its representation as a $2\times 2$ matrix accounts for the four dimensions.} we recognize that the larger collection of relative deformation-induced moir\'e is an exciting test bed to explore new moir\'e physics. A vdW heterostructure is said to be \emph{heterostrained} if its two lattices are under different strain states \citep{harley_disloc}. In this paper, we refer to the mutually exclusive twisted and heterostrained states using an umbrella term, \emph{heterodeformation}. The use of heterostrains to tune the electronic properties of materials and explore new quantum states is the goal of \emph{straintronics} \citep{miao2021straintronics}, an emerging research area.

Homostructures such as BGs, under a small twist ($1^\circ-3^\circ$) relative to the energetically favorable AB stacking, undergo atomic reconstruction due to spontaneous nucleation of interface dislocations, also referred to as \emph{strain solitons}. Recognizing the sensitivity of the electronic properties to atomic rearrangements, strain engineering offers an exciting route to modulate the electronic properties of BGs by controlling the dislocation network using strains \citep{Annevelink_2020,cazeaux2023relaxation,Cao2021, kim2022strain}. The overarching goal of this paper --- formulated to fully realize the potential of strain engineering for 2D heterostructures -- \emph{is to investigate and model atomic reconstruction in heterodeformed moir\'e superlattices}. Moreover, due to the large size of moir\'e superlattices, a high-throughput investigation of heterodeformations is computationally challenging, which motivates us to seek a continuum model for atomic reconstruction. In what follows,  we will identify the key features of atomic reconstruction observed in small-twist BG before formulating the objectives of this paper.

The atomic reconstruction \citep{Annevelink_2020,ZHANG_Tadmor_2017,ZHANG_Tadmor_2018,Lopes2012,Carr_2018_continuum,cazeaux2020energy,cazeaux2023relaxation,Gargiulo_2018,zhouVanWaalsBilayer2015} in a small-twist BG is a consequence of the interplay between interfacial energy and the elastic energies of the two lattices. The former is often described using the generalized stacking fault energy (GSFE) density, a periodic function of relative translations between the two AB-stacked lattices of the BG. The periodicity of the GSFE is derived from the bicrystallography of the interacting lattices. Under small twists relative to the AB stacking, the interfacial energy increases as the induced relative translations between the two lattices lead to regions of low-commensurability (high interfacial energy) interspersed with the highly commensurable AB stacking. The twisted BG responds to lower the interfacial energy by an atomic rearrangement that tends to increase (decrease) areas of high (low) commensurability. Due to the periodicity of the GSFE, the structural relaxation results in lines of displacement ``jumps'' that manifest as interface dislocation lines with the displacement ``jump'' as the Burgers vector \citep{alden2013strain,kumar2016limits}. Moreover, the Burgers vectors are parallel to the dislocation lines and their magnitude is equal to that of the smallest lattice vector of graphene. Therefore, the structural relaxation in a small-twist BG can be interpreted as elastic distortions associated with the formation of an array of screw lattice dislocations. Since the elastic energy diverges for discontinuous displacements, the ``jumps'' occur as localized displacement gradients, which implies the dislocation lines are diffused. The balance between the interfacial and elastic energies, which ultimately determines the network and the thickness of the dislocation lines, is at the core of the Frenkel--Kontorova continuum model for small-twist BGs. In this paper, the terms 'atomic relaxation' and 'structural relaxation' are used synonymously. 

Our study of atomic reconstruction in heterodeformed moir\'e is guided by the energetics of structural relaxation in small-twist BGs. We begin by hypothesizing that the structural relaxation of a heterodeformed moir\'e is also a consequence of interface dislocations and investigating the hypothesis using atomistic simulations. Instead of a bonafide 2D heterostructure, we use large-twist BG in our atomistic study due to the greater reliability of its interatomic potential, confirmed using Density Functional Theory (DFT). Moreover, it is reasonable to interpret a large-twist BG as a heterostructure since its lattices differ considerably. We show that the $21.786789^\circ$ large-twist BG, when subjected to heterostrains, results in strain localization in a network of lines, suggesting the formation of interface dislocations. Interestingly, the Burgers vector of the dislocations is smaller than that of the small-twist case.

In the presence of two distinct lattices, the notion of an interface dislocation has to be made precise as it is not clear to which lattice its Burgers vector belongs. The interpretation of a large-twist moir\'e as a network of lattice screw dislocations breaks down as the dislocation cores overlap. To resolve this ambiguity, we turn our attention to grain/phase boundaries. Similar to a small-twist BG, small-tilt angle grain boundaries can be interpreted as an array of lattice dislocations. For large tilt angles, however, a grain boundary dislocation is defined as a defect in the translation invariance of the boundary \citep{bollman_1974}. The translational invariance is derived by introducing two additional lattices --- coincident site lattice (CSL) and the displacement shift complete lattice (DSCL) --- originating from bicrystallography \citep{Balluffi_1982}. The CSL is the intersection of the two lattices, and the DSCL is the smallest lattice that contains the two lattices. In 2D heterostructures, it is straightforward to see that the CSL \emph{is} the moir\'e superlattice. On the other hand, the DSCL conveys the translational invariance of the interface --- displacing one lattice relative to the other by a DSCL vector preserves the structure of the interface. In other words, if a heterointerface hosts a dislocation, its Burgers vector must be a DSCL vector. While \citet{koda2016coincidence} have used the CSL to identify heterodeformed moir\'es, the use of DSCL to study interface dislocations remains largely unexplored.\footnote{
    A notable exception is the work of \citet{ishikawa2016interfacial} where the DSCL and the moir\'e superlattice are used to infer the atomic structure of twisted few-layer graphene, which is in the spirit of \emph{moir\'e metrology} \citep{annevelink2021moire}.
} One of the key highlights of this paper is the application of Smith Normal Form (SNF) bicrystallography to characterize interface dislocations. SNF bicrystallography is an algebraic framework developed by the last authors' group to explore bicrystallography properties such as the translational invariance \citep{ADMAL2022}. In particular, it informs us that the Burgers vector (smallest DSCL lattice vector) is inversely proportional and a rational multiple of a CSL vector.

Based on the atomistic simulations of heterodeformed BG and the SNF bicrystallography framework,  we build a generalized Frenkel--Kontorova (GFK) model. The generalization relative to the classical Frenkel--Kontorova model stems from key features of the GFK model --- frame-invariance and defect-free \emph{natural configurations}, which may include stackings that are not necessarily of the lowest energy.\footnote{For example, in addition to the AB-stacked BG, a large-twist BG with a twist angle of $21.786789^\circ$ is also a natural configuration.}
The GFK model generalizes the previous model of \citet{Koshino_2017} to large heterodeformations, including large twists. Unlike the model of \citet{Koshino_2017}, which was developed exclusively for infinite systems, the model describes finite systems as well, wherein configurational forces due to surface tension play an important role. 

This paper is organized as follows. In \sref{sec:atomistics}, we explore structural relaxation in a BG subjected to large heterodeformations using DFT-informed atomistic simulations and demonstrate the nucleation of interface dislocations. In \sref{sec:snf}, we review SNF bicrystallography and apply it to characterize the interface dislocations in heterodeformed BGs. In \sref{sec:continuum}, we build the GFK model and implement and validate it in \sref{sec:numerics}. We summarize and conclude in \sref{sec:conclusion}.

\noindent
\emph{Notation}:
We use lowercase bold letters to denote vectors, and uppercase bold letters to denote second-order
tensors, unless stated otherwise. The gradient, divergence, and curl operators are denoted by the symbols $\nabla$, $\divr$, and curl respectively. We use the symbol $\otimes$ to denote the tensor product of two vectors, and $\cdot$ to denote the inner product of two vectors or tensors.

\section{Atomic scale investigation of structural relaxation under large heterodeformations}
\label{sec:atomistics}
This section investigates the structural relaxation of 2D heterostructures using atomistic simulations of heterostrained BG, with the understanding that under large twists, a BG serves as a surrogate for a 2D heterostructure. The relaxation is restricted to being in-plane. Simulations are performed using Large-Scale Atomic/Molecular Massively Parallel Simulator (LAMMPS) \citep{plimpton1995fast}. Beginning with a small-twist BG, we systematically explore various small and large heterodeformations that result in atomic reconstruction. We will demonstrate that atomic reconstruction due to large heterodeformations results from interface dislocations, whose Burgers vector and the network are markedly different from those observed under small twists. We will revisit the examples of this section in \sref{sec:continuum} using a continuum model.

\begin{table}[t]
\centering
\small
\begin{tabular}{c|c c c c c c c c}
       &   $C [\si{\meV}]$ & $C_0 [\si{\meV}]$       & $C_2 [\si{\meV}]$ &     $C_4 [\si{\meV}]$  &   $A [\si{\meV}]$  & $\delta [\si{\angstrom}]$  &   $\lambda [\si{\per \angstrom}]$ &   $z_0 [\si{\angstrom}]$ \\
\hline
KC-1 & $\num{3.030}$ & $\num{15.71}$  & $\num{12.29}$ & $\num{4.933}$  & $\num{10.238}$ &  $\num{0.578}$ & $\num{3.629}$  &  $\num{3.34}$.\\
KC-2  & $\num{6.678908e-4}$ & $\num{21.847167}$ & $\num{12.060173}$ & $\num{4.711099}$ & $\num{12.660270}$ & $\num{0.7718101}$ &  $\num{3.143921}$ & $\num{3.328819}$.
\end{tabular}
\caption{Two parameterizations of the KC potential.}
\label{table:parameters}
\end{table}
The simulated BGs are oriented such that the normal to the lattices is along the $X_3$ direction. Since we allow only in-plane relaxation, the distance between the two graphene lattices is held fixed during the simulation. The intralayer bonding in each graphene sheet is modeled using the reactive empirical bond order (REBO) potential \citep{Rebo_Brenner_2002}. The interlayer vdW interaction is described using the registry-dependent Kolmogorov--Crespi (KC) potential \citep{kolmogorov2005kc}. We investigate structural relaxation for two parametrizations of the KC potential, denoted as KC-1 \citep{kolmogorov2005kc} and KC-2 \citep{ouyang2018nanoserpents} with parameters listed in Table~\ref{table:parameters}. 
Since the KC-2 model was developed for BG systems under an out-of-plane compression, it may be viewed as an improvement of the KC-1 model. We will, however, explore both KC-1 \emph{and} KC-2 models while investigating large heterodeformations as the qualitative differences in the respective GSFEs lead to markedly different structural relaxations. 

Periodic boundary conditions (PBCs) are imposed along two in-plane directions to avoid the influence of free boundary lines. Since PBCs necessitate the existence of a periodic supercell, we are restricted to bilayer configurations wherein the intersection of the projections (on the $X_1-X_2$ plane) of the two lattices is a 2D superlattice. In other words, PBCs can be enforced if and only a CSL exists. The process of identifying heterodeformations that admit PBCs can be formalized as follows. Two 2D (multi) lattices $\mathcal A$ and $\mathcal B$, with structure matrices\footnote{The two basis vectors of a lattice are stored as columns of its structure matrix.} $\bm A$ and $\bm B$, respectively, are coincident on a 2D CSL if and only if $\bm T:= \bm A^{-1}\bm B$ is a rational matrix. We refer to $\bm T$ as the \emph{transition} matrix. In a homostructure, if lattice $\mathcal B$ is obtained by deforming lattice $\mathcal A$ using a deformation gradient $\bm F$, then $\bm B=\bm F \bm A$, and all deformations that result in a rational $\bm A^{-1}\bm F \bm A$ yield a CSL, and therefore, amenable to PBCs. In \sref{sec:snf}, we will show that the bicrystallographic properties of a heterodeformed moir\'e can be deduced from the algebraic properties of the transition matrix. For example, the ratios of the areas of the primitive unit cells ---
\begin{equation}
    \Sigma_{\mathcal A} = \frac{\text{Area(CSL)}}{\text{Area}(\mathcal A)},\quad
    \Sigma_{\mathcal B} = \frac{\text{Area(CSL)}}{\text{Area}(\mathcal B)},
    \label{eqn:sigma}
\end{equation}
are always integers, and if $\mathcal A$ and $\mathcal B$ have the same density, $\Sigma_{\mathcal A}= \Sigma_{\mathcal B}=:\Sigma$. The two basis vectors of the CSL of a heterodeformed moir\'e are chosen as the in-plane simulation box vectors. Therefore, the number of simulated atoms is equal to $n_{\mathcal A} \Sigma_{\mathcal A}+n_{\mathcal B} \Sigma_{\mathcal B}$, where the factors $n_{\mathcal A}$ and $n_{\mathcal B}$ represents the number of basis atoms in the primitive unit cells of the respective 2D multilattices. In all our simulations, $\mathcal A$ is a graphene lattice formed by the structure matrix
\begin{equation*}
    \bm A = \frac{a}{2} \begin{bmatrix}
    0 &  -\sqrt{3}\\
    2 & -1
    \end{bmatrix},
\end{equation*}
and $\mathcal B$ is a deformation or rotation of $\mathcal A$, and placed at a prescribed interplanar distance in the $X_3$ direction from $\mathcal A$. The heterodeformed configurations studied in this paper are calculated using an algorithm (see Algorithm~2 in \citet{ADMAL2022}) derived from \thref{thm:F} in  \ref{sec:append_Coinciden_rel}, which generates heterostrained moir\'es of various sizes and strains within prescribed upper bounds.

Atomic reconstruction is simulated by minimizing the total energy with respect to in-plane displacements of atoms using the fast inertial relaxation engine (FIRE) algorithm \citep{bitzek2006fire} with an energy tolerance  and force tolerance of $\SI{1e-20}{\eV}$ and $\SI{1e-20}{\eV \per \angstrom}$, respectively. The resulting displacements of atoms are analyzed to interpret them in terms of interface dislocations. 

\subsection{Atomic reconstruction in a BG under a small twist and a small strain}
\label{sec:atomistics_small}
In this section, we present simulations of atomic reconstruction in a BG under two small heterodeformations --- a) a $0.2992634^\circ$ twist and b) a pure stretch of
\begin{equation}
    \bm U = 
    \begin{bmatrix}
        1.004219 &  0\\
        0        &  0.995781
    \end{bmatrix}
    \label{eqn:U_smallTwist}
\end{equation}
relative to the AB-stacked $\Sigma 1$ configuration. Since the AB-stacked configuration corresponds to $\bm F= \bm R(60^\circ)$, lattice $\mathcal B$ of the $0.2992634^\circ$-twisted BG is constructed using $\bm F= \bm R(60.2992634^\circ)$, and for the heterostrained case, $\bm F= \bm R(60^\circ)\bm U$. The interplanar distance is fixed at $\SI{3.34}{\angstrom}$. The basis vectors of the corresponding CSLs,  
\begin{subequations}
    \begin{align}
        \text{twist: } \bm b_1 = 470.824979 \, \bm e_1, 
        &\quad 
        \bm b_2=235.412488 \,  \bm e_1 +407.746391\, \bm e_2, \text{ and }\\
        \text{heterostrain: } \bm b_1 = -581.794\, \bm e_1, 
        &\quad 
        \bm b_2=-287.199\,  \bm e_1 + 505.965\, \bm e_2,
    \end{align}
    \label{eqn:pbc_small_strain}
\end{subequations}
define the  respective periodic boxes of the simulations. From \eqref{eqn:sigma} and \eqref{eqn:pbc_small_strain}, it follows that $\Sigma=\num{36631}$ for the $0.2992634^\circ$-twisted BG, whereas $\Sigma_{\mathcal A}=\num{56168}$ and $\Sigma_{\mathcal B}=\num{56169}$ for the heterostrained BG.

\begin{figure}[H]
    \centering
    \subfloat[]
    {
        \includegraphics[height=0.3\textwidth]{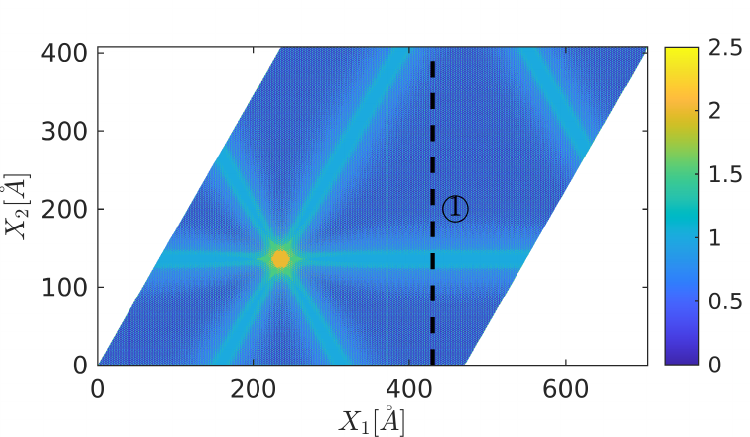}
        \label{fig:glob_tw_en_den_lammps}
    }
    \subfloat[]
    {
        \includegraphics[height=0.28\textwidth]{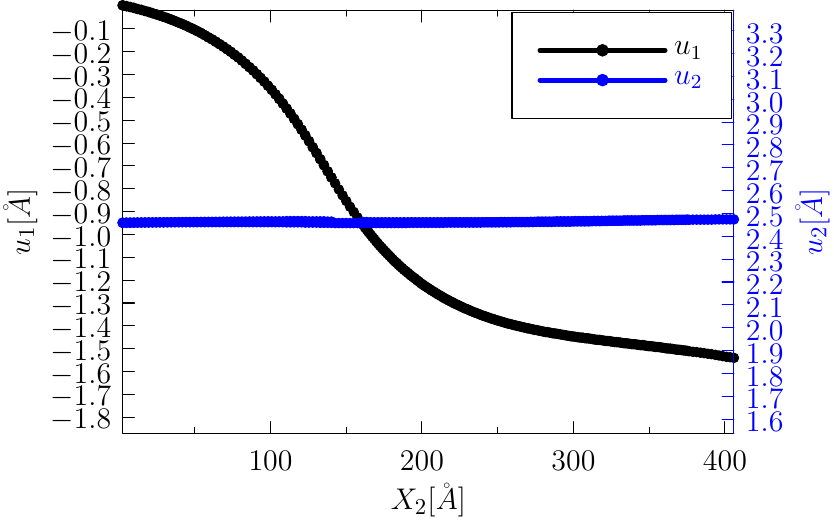}
        \label{fig:s1_twist_u_y}
    }
    \\
    \subfloat[]
    {
        \includegraphics[height=0.3\textwidth]{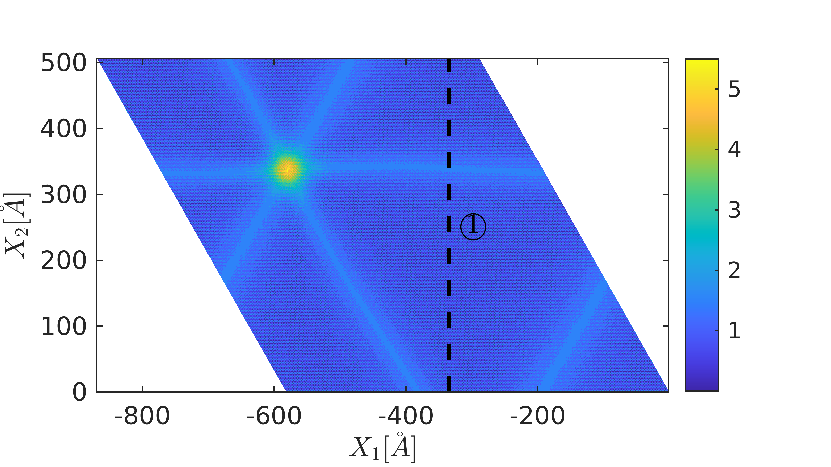}
        \label{fig:glob_str_en_den_lammps}
    }
    \subfloat[]
    {
        \includegraphics[height=0.28\textwidth]{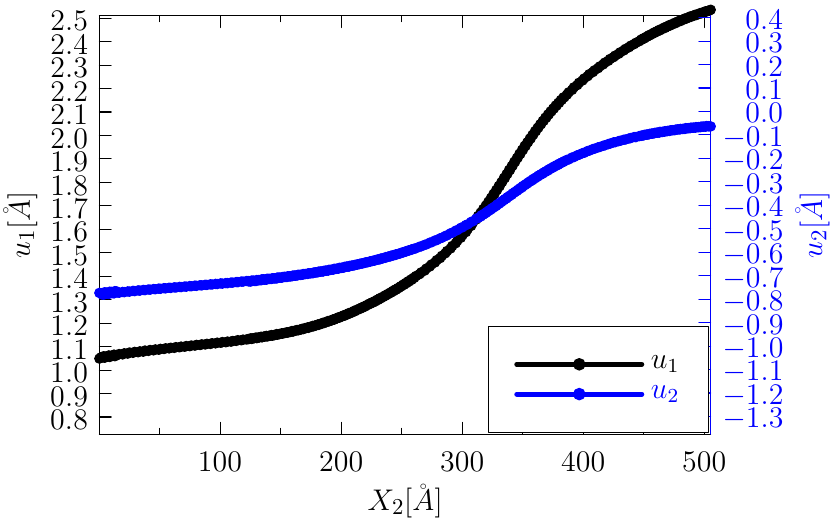}
        \label{fig:s1_stretch_u_y}
    } 
    \caption{Atomic reconstruction in a BG under a small twist (top row) and a small heterostrain (bottom row). \psubref{fig:glob_tw_en_den_lammps}, \psubref{fig:glob_str_en_den_lammps} Plots of atomic energy density [$\si{\meV \per \angstrom\squared}$] show a triangular network of interface dislocations. The dislocation lines separate triangular domains of low-energy AB-stacking. \psubref{fig:s1_twist_u_y}, \psubref{fig:s1_stretch_u_y} Line plots of the displacement components $u_1$ and $u_2$, measured along the dashed lines in \psubref{fig:glob_tw_en_den_lammps} and \psubref{fig:glob_str_en_den_lammps}. The displacements are measured relative to the untwisted AB-stacked configuration.}
    \label{fig:glob_tw_lammps}
\end{figure}
The color density plots of atomic energy density shown in \frefs{fig:glob_tw_en_den_lammps}{fig:glob_str_en_den_lammps} highlight the triangular dislocation network in the twisted and strained BGs, respectively. The high-energy nodal regions correspond to the AA stacking, and the interiors of the triangular domains are in AB stacking. \frefs{fig:s1_twist_u_y}{fig:s1_stretch_u_y} show line plots of displacements of atoms along the dashed lines in the respective energy density plots. The displacements are measured relative to the AB-stacking. Since \fref{fig:s1_twist_u_y} shows negligible displacement perpendicular to the dislocation line, interface dislocations in the twisted BG have a screw character. On the other hand, \fref{fig:s1_stretch_u_y} suggests the interface dislocations in the heterostrained case have a mixed character. In both cases, the Burgers vector magnitude (size of the displacement jump) is $<\SI{2.46}{\angstrom}$, the lattice constant of graphene, which implies the dislocations are not full dislocations. \citet{Annevelink_2020,harley_disloc} demonstrated that the partial dislocations have a pure edge character under a small biaxial heterostrain relative to the AB stacking.

\begin{figure}[H]
\centering
\includegraphics[width=0.48\textwidth]{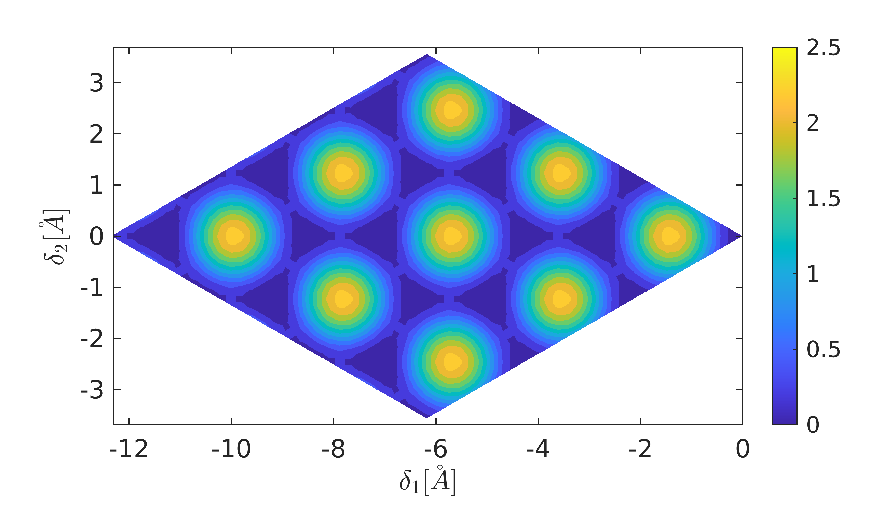}
\caption{GSFE of AB stacking in $\si{\milli \eV \per \angstrom\squared}$. The minima and maxima correspond to the AB and AA stackings, respectively. Parametrizations KC-1 and KC-2 yield nearly identical GSFEs.}
\label{fig:AB_gsfe}
\end{figure}
The origin of interface dislocations and their network pattern can be traced to the properties of AB-stacking's GSFE, shown in \fref{fig:AB_gsfe}. The GSFE of a BG configuration is a function of the relative displacement between the two layers. The GSFE of AB stacking is periodic with respect to the lattice vectors of graphene. Under small uniform heterodeformations, the relative displacement between the two layers is spatially varying, and therefore, the interfacial energy is sampled from various regions of the GSFE, including the maxima and the minima. The BG responds by an atomic rearrangement to increase (decrease) regions of AB (AA) stacking, which corresponds to minima (maxima) in the GSFE plot,
resulting in a juxtaposition of AB-stacked regions separated by dislocation lines. The  Burgers vector of a dislocation line separating two AB-stacked regions is the relative vector, with magnitude $\SI{1.42}{\angstrom}$, that connects the corresponding minima in the GSFE.\footnote{\fref{fig:s1_twist_u_y} does not quite recover the entire Burgers vector magnitude of $\SI{1.42}{\angstrom}$ as it includes displacements associated with elastic relaxation. In \sref{sec:continuum}, where we present our continuum model, we will discuss the procedure to accurately measure the Burgers vector from the displacement field.\label{fn:bg}} Moreover, the triangular network of dislocation lines with every AB-stacked region surrounded by three similar regions originates from the observation that each minimum in the GSFE is surrounded by three nearest minima. 

The arguments that helped us deduce the properties of dislocations from the GSFE are applicable only under small heterodeformations relative to the AB stacking. Under large heterodeformations, it is not clear if a heterostructure undergoes atomic reconstruction. If reconstruction occurs, its interpretation in terms of full/partial lattice dislocations breaks down as the dislocation cores overlap.\footnote{As the heterodeformation is measured relative to the AB stacking, the density of interface dislocations increases resulting in dislocation core overlap.}

\subsection{Atomic reconstruction in a BG under large heterodeformations}
\label{sec:atomistics_large}
\begin{figure}[H]
\centering
\subfloat[]
{
    \includegraphics[width=0.5\columnwidth,valign=c]{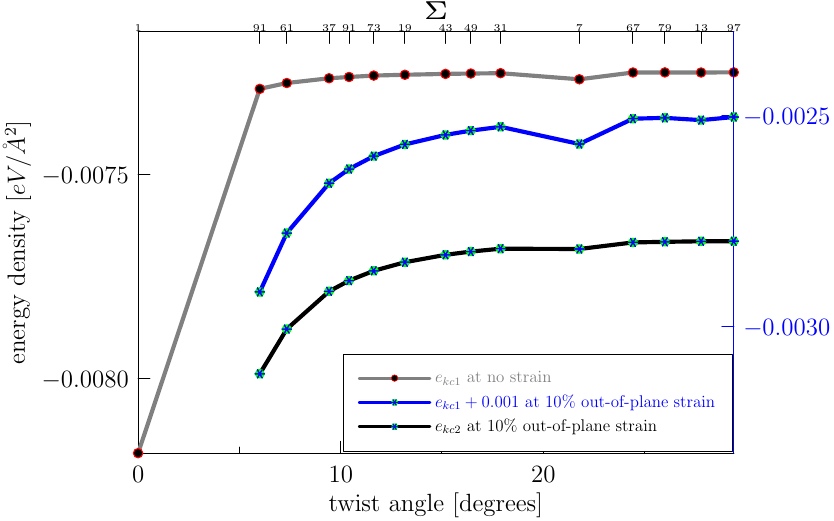}
    \label{fig:mis_vs_en}
}
    \subfloat[]
{
    \includegraphics[width=0.3\textwidth,valign=c]{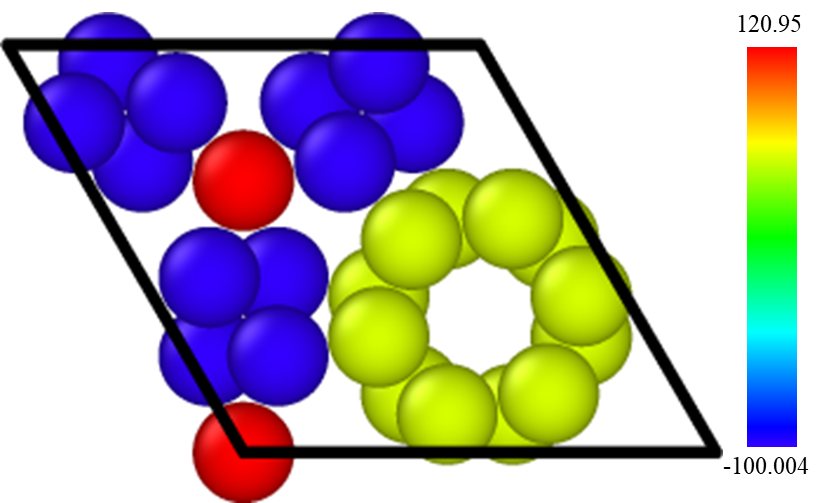}
    \vphantom
    {
        \includegraphics[width=0.5\columnwidth,valign=c]{figures/asy/mis_en.pdf}
    }
    \label{fig:21_78_stacking}
}
\caption{\psubref{fig:mis_vs_en} A comparison of interfacial energy density versus the twist angle of a BG under zero strain with that of a BG under a $10\%$ out-of-plane compression. \psubref{fig:21_78_stacking} Atomic energy density $[$\si{\meV \per \angstrom\squared}$]$ within a fundamental unit cell of the $\Sigma 7$ configuration, computed using the KC-1 potential.}
\end{figure}
It is well known that as the twist angle of a BG increases beyond a few degrees, the vdW interaction between the two lattices weakens, resulting in negligible structural relaxation \citep{Annevelink_2020,morovati2022interlayer}. Signatures of the presence or absence of interface dislocations can also be found in the plot of interface energy versus the twist angle, shown in Fig. \ref{fig:mis_vs_en}. For small twist angles relative to the AB stacking, the interface energy variation (grey plot) is non-convex --- a signature for potential defect nucleation. In contrast, for large twist angles, the interfacial energy is insensitive to the twist angle, and this justifies the absence of atomic reconstruction. While Fig. \ref{fig:mis_vs_en} explores only twists --- as opposed to heterodeformations --- it is plausible that atomic reconstruction does not occur in heterodeformed configurations far from the AB and AA stackings. However, by decreasing the interlayer distance using external pressure, the interlayer electronic coupling can be made to persist for certain large twist angles. Indeed, this strategy follows the historical trend of using external pressure to probe correlated electron physics. For example, the role of interlayer compression on the atomic reconstruction of 2D heterostructures has been shown to have a substantial influence on the band structure \citep{Carr2018, Chittari_2019, Das_2016_compression}. Moreover, experiments and first principle calculations \citep{nano2022localmoire,cheng2023moire} have shown that electronic scale effects are modified in large-twist bilayer graphene under out-of-plane compression.
The blue plot in \fref{fig:mis_vs_en} shows that under a $10\%$ out-of-plane strain, the interfacial energy from the KC-1 model is sensitive to twist angles in the neighborhood of $21.786789^\circ$. We will refer to the $21.786789^\circ$-twist BG as the $\Sigma 7$ configuration. More interestingly, the variation is non-convex in a neighborhood of $\Sigma 7$ twist. \fref{fig:21_78_stacking} shows the atomic energy densities of the $28$ atoms in the unit cell of the $\Sigma 7$ configuration.

The non-convexity of the interfacial energy in the neighborhoods of $\Sigma 1$ and $\Sigma 7$ configurations motivates us to hypothesize that atomic reconstruction occurs for small heterodeformations relative to the $\Sigma 7$ configuration in the presence of an out-of-plane compression. To investigate our hypothesis, we simulate the following two heterodeformations relative to the $\Sigma 7$ configuration --- a) a $0.170076^\circ$ twist and b) a pure stretch of
\begin{equation}
    \bm U = 
    \begin{bmatrix}
        1.010589 &  0\\
        0        &  0.997163
    \end{bmatrix}.
    \label{eqn:U_largeTwist}
\end{equation}
In other words, the heterodeformations are given by $\bm F=\bm R(81.956865^\circ)$ for the former, and $\bm F=\bm R(81.786789^\circ) \bm U$ for the latter case. The box vectors in the two simulations are 
\begin{subequations}
    \begin{align}
        \text{twist: } \bm b_1 = 313.233 \, \bm e_1, 
        &\quad 
        \bm b_2=156.616 \,  \bm e_1 + 271.267\, \bm e_2, \text{ and }\\
        \text{heterostrain: } \bm b_1 = 320.679\, \bm e_1, 
        &\quad 
        \bm b_2=1.04735\,  \bm e_1 + 548.127\, \bm e_2,
    \end{align}
    \label{eqn:pbc_large_strain}
\end{subequations}
which imply  $\Sigma=16213$, and $\Sigma_{\mathcal A}=33539$ and $\Sigma_{\mathcal B}=33282$, respectively.

\begin{figure}[H]
    \centering
        \subfloat[]{
        \includegraphics[width=0.4\textwidth]{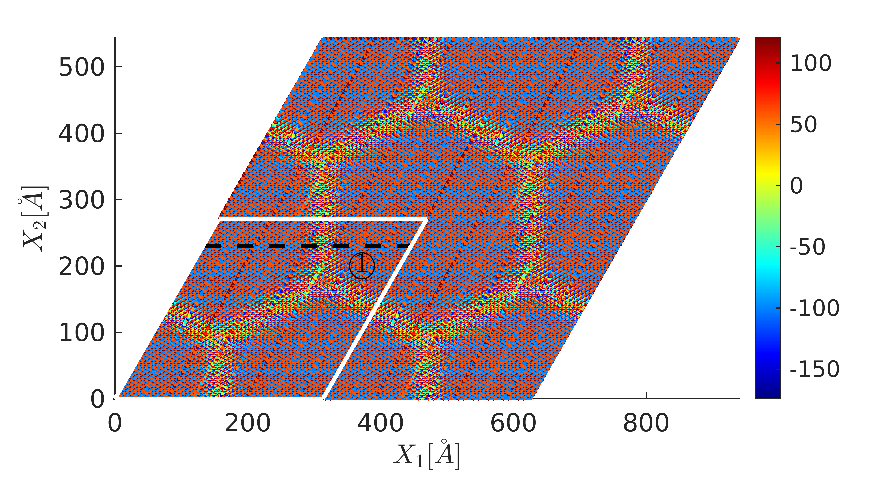}
        \label{fig:twist_local_old_res}
        }
        \subfloat[]{
        \includegraphics[width=0.35\textwidth]{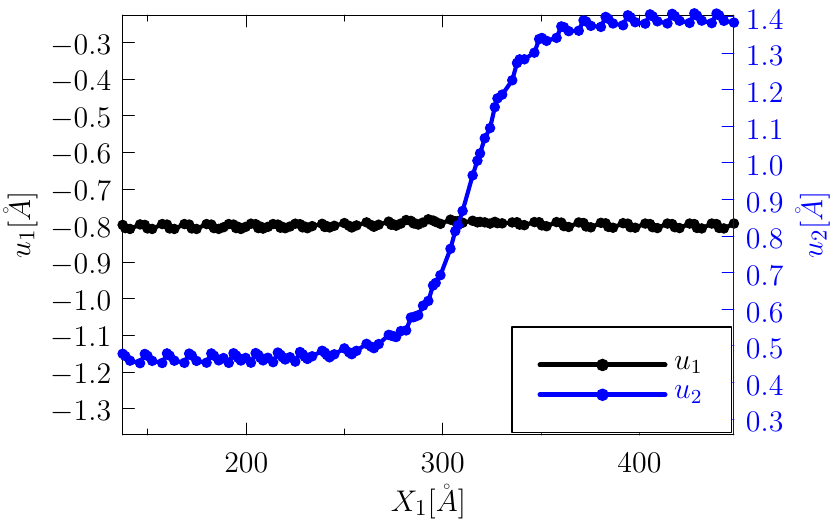}
        \label{fig:twist_x_ux_old}
        }
        \\
        \subfloat[]{
        \includegraphics[width=0.4\textwidth]{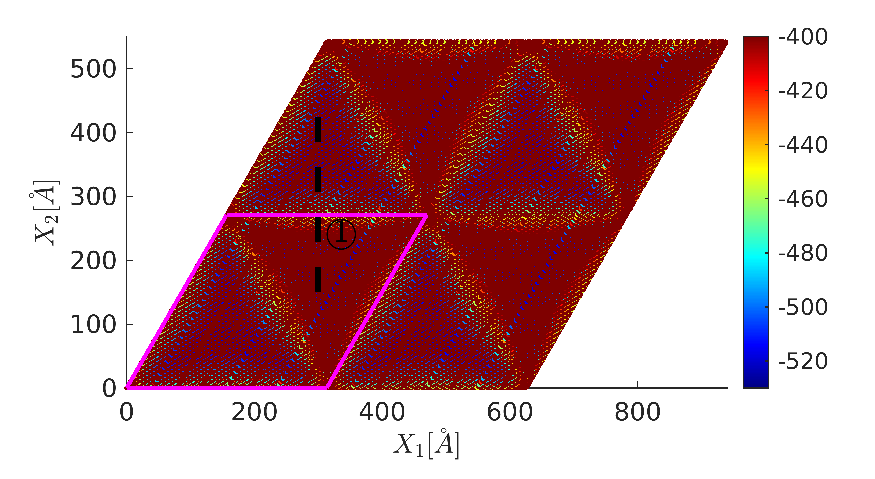}
        \label{fig:twist_local_new_res}
        }
        \subfloat[]{
        \includegraphics[width=0.35\textwidth]{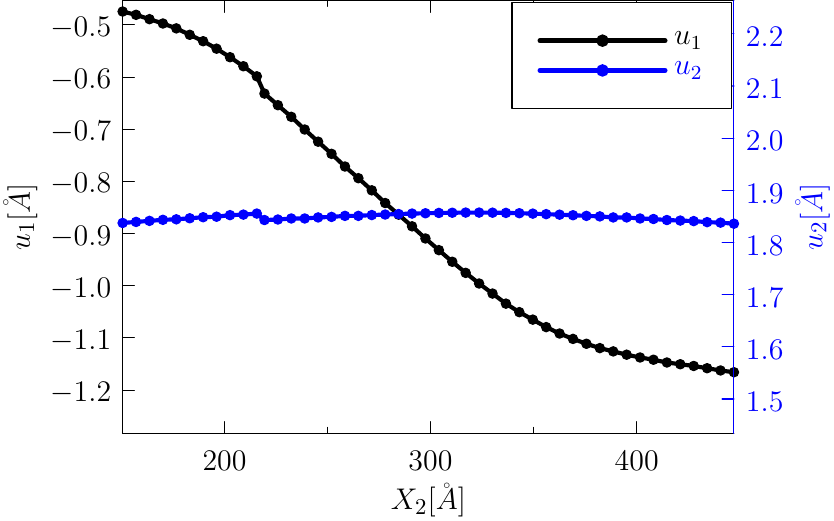}
        \label{fig:twist_x_ux_new}
        }
    \caption{Atomic reconstruction in a large-twist ($21.786789^\circ+0.170076^\circ$) BG using the KC-1 (top row) and the KC-2 (bottom row) models . \psubref{fig:twist_local_old_res}, \psubref{fig:twist_local_new_res} Plots of atomic energy density [$\si{\meV \per \angstrom\squared}$] show a honeycomb and a triangular network of interface dislocations. \psubref{fig:twist_x_ux_old}, \psubref{fig:twist_x_ux_new} Line plots of the displacement components $u_1$ and $u_2$, measured along the dashed lines in \psubref{fig:twist_local_old_res} and \psubref{fig:twist_local_new_res}. The displacements are measured relative to the untwisted $\Sigma 7$ configuration.}
    \label{fig:local_twist}
\end{figure}

\fref{fig:local_twist} shows results of the large-twist simulations using the KC-1 (\frefs{fig:twist_local_old_res}{fig:twist_x_ux_old}) and KC-2 (\frefs{fig:twist_local_new_res}{fig:twist_x_ux_new}) models. Comparing the color density plots of atomic energy density in \frefs{fig:twist_local_old_res}{fig:twist_local_new_res}, we note that the KC-1 model yields a honeycomb interface dislocation network, while KC-2 results in a triangular network, similar to the small twist case (\fref{fig:glob_tw_en_den_lammps}). The atomic arrangement in the interiors of the triangular and hexagonal domains is that of the $\Sigma 7$ stacking.
The plots in \frefs{fig:twist_local_old_res}{fig:twist_local_new_res} are of the atomic energy density relative to the total energy density of the $\Sigma 7$ stacking.  Therefore, we expect the energy density in the interior of the triangular/hexagonal domains to be zero. \frefs{fig:twist_local_old_res}{fig:twist_local_new_res}, however, do not reflect this due to the variation in the energy densities of the $28$ atoms in the primitive unit cell of the $\Sigma 7$.\footnote{In other words, if the energy densities in \frefs{fig:twist_local_old_res}{fig:twist_local_new_res} were spatially averaged using a weighting function with a $\Sigma 7$ unit cell-shaped averaging domain, the resulting fields will be zero in the domain interiors.}

To identify the nature of the interface dislocations, we plotted (\frefs{fig:twist_x_ux_old}{fig:twist_x_ux_new}) the displacements of atoms along the dashed lines in \frefs{fig:twist_local_old_res}{fig:twist_local_new_res}. 
The displacements are measured relative to the untwisted $\Sigma 7$ stacking. The displacement line plots show negligible displacement perpendicular to the vertical dislocation line, suggesting a screw character. Moreover, the displacement ``jumps'' suggest the Burgers vector magnitude is $\ll$ than that of the partial dislocations, noted in \sref{sec:atomistics_small}. Interestingly, the Burgers vector of dislocations from the KC-1 model is larger than those from the KC-2 model.
\begin{figure}[t]
\centering
\includegraphics[width=0.45\columnwidth]{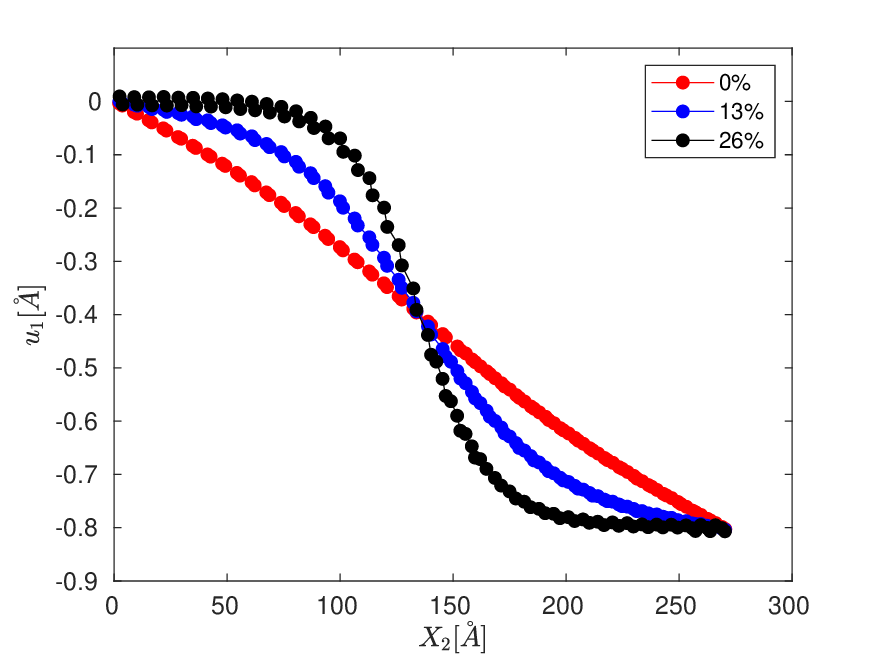}
\caption{Burgers Vector for $21.786789^\circ+0.170076^\circ$ twisted BG at different magnitudes of out-of-plane compression calculated using the KC-1 parametrization}
\label{fig:compression}
\end{figure}
For larger out-of-plane compression, the non-convexity of the interfacial energy increases in the neighborhood of $\Sigma 7$. Therefore, we expect sharper displacement ``jumps'', which is confirmed in \fref{fig:compression}.

\begin{figure}[H]
    \centering
    \subfloat[]
    {
        \includegraphics[height=0.35\textwidth]{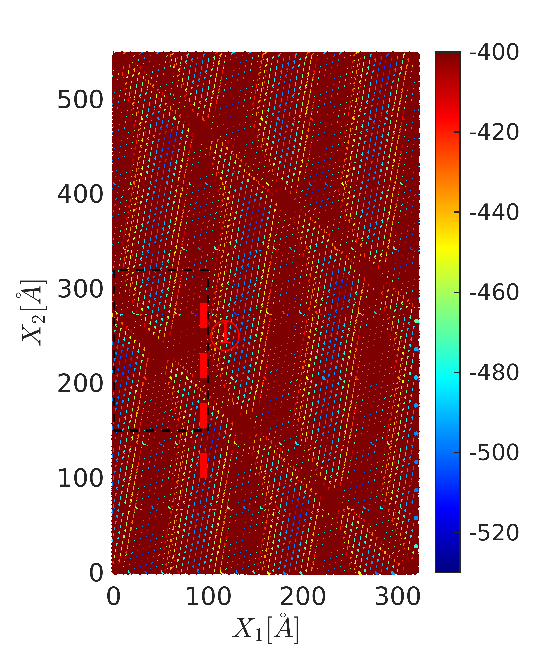}
        \label{fig:local_stretch_en_den_lammps_new}
    }
    \subfloat[]
    {
        \includegraphics[height=0.35\textwidth]{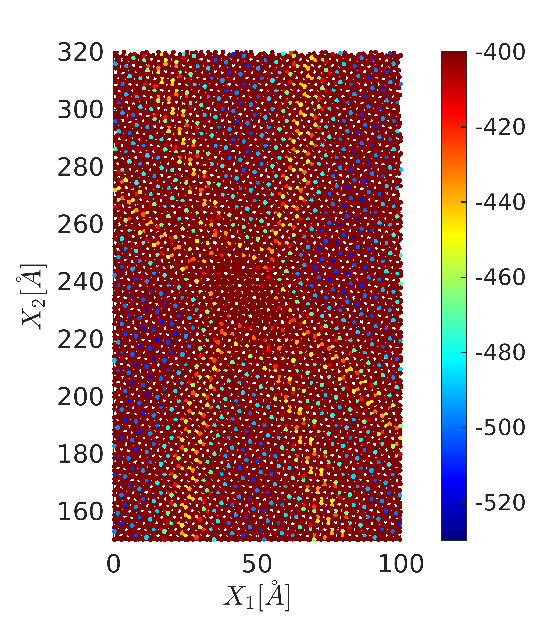}
        \label{fig:local_stretch_en_den_lammps_magnify}
    }
    \subfloat[]
    {
        \includegraphics[height=0.25\textwidth]{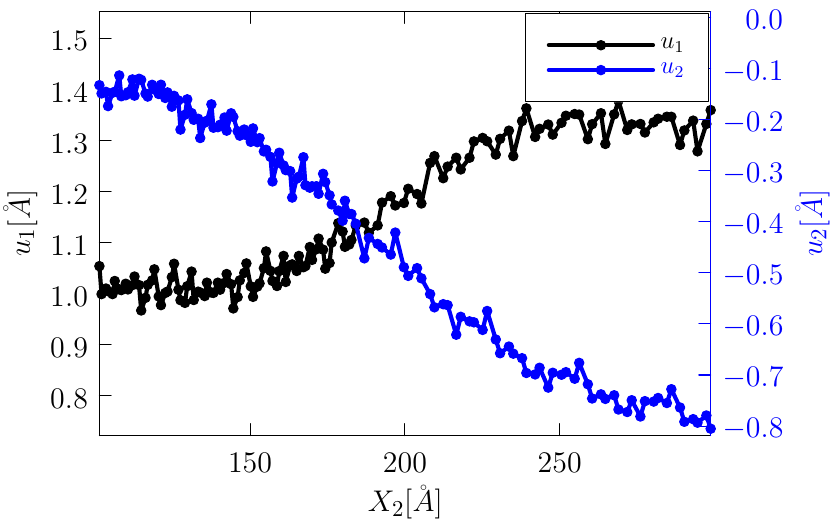}
        \label{fig:local_hetero_burgers_u_lammps_new}
    }
    \caption{Atomic reconstruction under a large heterodeformation.  \psubref{fig:local_stretch_en_den_lammps_new} Color density plot of atomic energy density [$\si{\meV \per \angstrom\squared}$] highlighting the network of interface dislocations in a BG heterodeformed relative to the $\Sigma 7$ configuration using KC-2 parametrization. \psubref{fig:local_stretch_en_den_lammps_magnify} A magnified view of the dislocation network in \psubref{fig:local_stretch_en_den_lammps_new}. \psubref{fig:local_hetero_burgers_u_lammps_new} shows variations of the displacement components along the dashed line $1$, shown in \psubref{fig:local_stretch_en_den_lammps_new}. The displacements are measured relative to the $\Sigma 7$ configuration.}
    \label{fig:hetero_local_lammps}
\end{figure}
\fref{fig:hetero_local_lammps} shows simulation results of a BG under large heterodeformation. Similar to the large twist case, we observe interface dislocations that form a distorted triangular network surrounding regions of $\Sigma 7$ stackings. In addition, the displacement line plots along the dashed line in  \fref{fig:local_stretch_en_den_lammps_new} suggest the dislocations have both screw and edge components, similar to heterostrained BG in Fig. \ref{fig:glob_str_en_den_lammps}.

Summarizing, this section conclusively demonstrates that atomic reconstruction occurs when a $\Sigma 7$ BG is subjected to small heterodeformations.\footnote{Note that while the heterodeformation is small relative to the $\Sigma 7$ configuration, it is large relative to the AB stacking.} Analogous to the AB-stacking, the $\Sigma 7$ configuration is energetically favorable and defect-free. When the $\Sigma 7$ configuration is subjected to a small heterodeformation, atomic reconstruction ensues through strain localization along a network of lines, which we interpret as dislocations. However, we are yet to identify the crystallographic origin of the observed dislocations and their relatively short Burgers vector. In the next section, we will present SNF bicrystallography, an algebraic framework to study the geometric properties of moir\'e superlattices. In particular, we will apply SNF to arrive at a rigorous definition for interface dislocations that is applicable across all heterodeformations.

\section{Bicrystallography and interface dislocations}
\label{sec:snf}
The goal of this section is to define interface dislocations in 2D heterostructures, including homostructures, under large heterodeformations. An interface dislocation is a line defect that breaks the \emph{translational invariance} of a \emph{defect-free interface}. Low-energy configurations, such as $\Sigma 1$ and $\Sigma 7$ interfaces in a twisted BG, are considered defect-free. In what follows, we will describe a framework to characterize the translational invariance of defect-free interfaces, which ultimately yields the set of interface dislocations a heterointerface can host. Recall from \sref{sec:atomistics_small} that the Burgers vector of interface dislocations, formed due to a small heterodeformation of the AB stacking, originates from the GSFE of the AB-stacking. In particular, the periodicity of the GSFE conveys the translational invariance of the interface. This motivates us to investigate the GSFE of the $\Sigma 7$ interface. We use SNF bicrystallography to identify the periodicity of GSFE of a defect-free heterointerface and use it to identify interface dislocations in the $\Sigma 7$ configuration. In what follows, we describe the main results of SNF bicrystallography. For further details, we refer the reader to \aref{sec:snf_details} and \citet{ADMAL2022}.

SNF bicrystallography is a framework that utilizes the Smith normal form for integer matrices to analyze the bicrystallography of crystal interfaces. It was developed by \citet{ADMAL2022} to enumerate disconnections (dislocations with a step height) in grain boundaries of simple lattices. Due to the generality of the SNF framework, it is a powerful tool to analyze defects in heterostructures. The open Interface Lab (\texttt{oILAB}), a C++ dimension-independent implementation of SNF bicrystallography archived at   \url{https://github.com/oilab-project/oILAB.git}, was used to generate the heterostructures studied in this paper. The framework begins with two lattices $\mathcal A$ and $\mathcal B$ --- with respective structure matrices $\bm A$ and $\bm B$ --- that overlap on a lattice $\mathcal C:= \mathcal A \cap \mathcal B$, called the \emph{coincident site lattice}. In the context of this paper, $\mathcal A$ and $\mathcal B$ are 2D multilattices, and $\mathcal C$ is a moir\'e superlattice, which we will assume to be 2D as well. A key step in SNF bicrystallography is the transformation of basis vectors of lattices such that the new basis vectors of the lattices, collected in structure matrices $\Ap$, $\Bp$, and $\Cp$, are parallel. This transformation makes it straightforward to introduce a fourth lattice $\mathcal D$, called the \emph{displacement shift complete lattice} (DSCL), defined as the smallest lattice that contains $\mathcal A$ and $\mathcal B$, and therefore $\mathcal C$ as well.\footnote{$\mathcal D$ is a fictitious lattice as some of its points are unoccupied.} Interestingly, the integer algebra of SNF reveals that the ratios of the four lattices are integers, given in terms of integers $\Sigma_{\mathcal A}$ and $\Sigma_{\mathcal B}$ (defined in \eqref{eqn:sigma}), and are related as
\begin{equation}
    \det(\bm C) \, \det(\bm D) =  \det(\bm A) \, \det(\bm B).
    \label{eqn:ABCD_relation}
\end{equation}
\begin{figure}[htbp]
    \centering
    \subfloat[]
    {
        \begin{tikzpicture}[scale=1]
            \node[]  at (0,1.5) {\frame{
            \includegraphics[width=0.48\textwidth,valign=t]{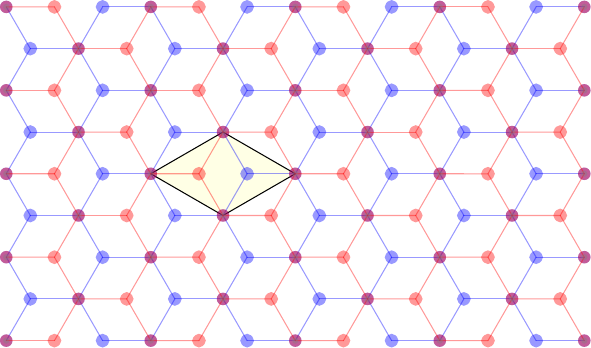}
            }};
            \fill[draw=white!60!blue,fill=white,line width=0.5mm] (-0.15\columnwidth,-1.0) circle (2.5pt); 
            \fill[fill=white!60!blue] (-0.15\columnwidth,-1.3) circle (3pt) node[anchor=south west]{\scriptsize{$\mathcal{A}$ lattice}};
           \fill[draw=white!60!red,fill=white,line width=0.5mm] (-0.05 \columnwidth,-1.0) circle (3pt);
           \fill[fill=white!60!red] (-0.05 \columnwidth,-1.3) circle (3pt) node[anchor=south west]{\scriptsize{$\mathcal{B}$ lattice}};
           \fill[draw=blue!50!red,opacity=0.6,fill=white,line width=0.5mm] (0.05\columnwidth,-1.0) circle (3pt); 
           \fill[fill=blue!50!red,opacity=0.6] (0.05\columnwidth,-1.3) circle (3pt) node[anchor=south west]{\scriptsize{$\mathcal C$ CSL}};
           \fill[fill=black!60!white,opacity=0.6] (0.15\columnwidth,-1.0) circle (1pt) node[right]{\scriptsize{$\mathcal D$ DSCL}};
        \end{tikzpicture}
        \label{fig:0_schem}
    }
    \subfloat[]
    {
        \begin{tikzpicture}[scale=1]
            \node[]  at (0,0) {\frame{
            \includegraphics[width=0.48\textwidth]{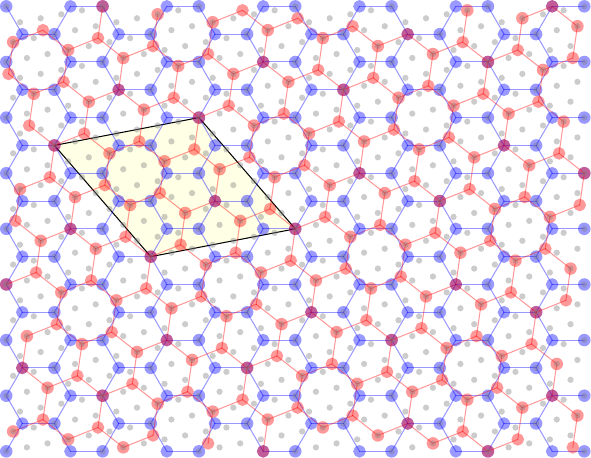}
            }};
        \end{tikzpicture}
        \label{fig:81_schem}
    }
    \caption{\psubref{fig:0_schem} AB-stacked ($\SI{0}{\degree}$ twist) bilayer graphene forming a $\Sigma=1$ moir\'e. \psubref{fig:81_schem} $21.786789^\circ$ twisted BG resulting in a $\Sigma=7$ moir\'e. Open circles represent the second basis atom of graphene. The two graphene lattices, $\mathcal A$ and $\mathcal B$, are shown in blue and red, while the moir\'e superlattice ($\mathcal C$) and the DSCL ($\mathcal D$) are shown in purple and gray, respectively. The highlighted region represents a unit cell of $\mathcal C$. The ratio of the unit cell size of the moir\'e superlattice to that of the graphene lattice and the corresponding ratio between the graphene lattice and $\mathcal D$ are equal to $\Sigma$.}
    \label{fig:schem}
\end{figure}
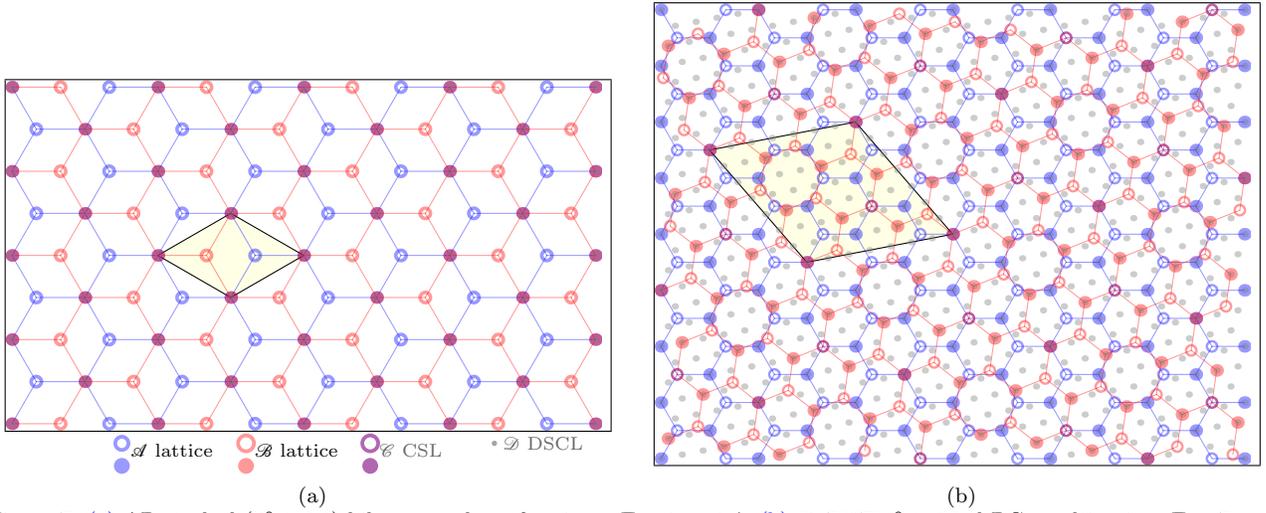
\frefs{fig:0_schem}{fig:81_schem} show the four lattices in the AB-stacked and the $\Sigma 7$ BG configurations, respectively. The two graphene multilattices $\mathcal A$ and $\mathcal B$ are shown in red and blue. The CSL $\mathcal C$ is marked in purple (red+blue), and the DSCL $\mathcal D$ in grey. Since $\Sigma=1$ in the AB-stacked configuration, the four lattices (not including the basis atoms of $\mathcal A$ and $\mathcal B$) coincide in \fref{fig:0_schem}. On the other hand, the DSCL (CSL) in \fref{fig:81_schem} is $7$ times smaller (larger) than the graphene lattice.

The following theorem characterizes the translational invariance of the interface and highlights the importance of the DSCL. The parallel bases vectors resulting from SNF play a critical role in the proof (see \citet{ADMAL2022}) of the theorem. 
\begin{theorem}
\label{thm:csl_shift}
Translating lattice $\mathcal A$ by a vector $\bm b \in \mathcal D$ with $\mathcal B$ fixed results in a shift $\bm \lambda_{\mathcal A} \in \mathcal B$ of the CSL. In other words
\begin{equation*}
    (\mathcal A + \bm b) \cap \mathcal B = \mathcal C + \bm \lambda_{\mathcal A}.
\end{equation*}
In addition, the shift is linear in $\bm b$, i.e. 
\begin{equation}
\label{eqn:linear}
\bm \lambda_{\mathcal A} (\bm b_1 + \bm b_2) = 
\bm \lambda_{\mathcal A} (\bm b_1) +
\bm \lambda_{\mathcal A} (\bm b_2), \text{ for any $\bm b_1,\bm b_2 \in \mathcal D$}.
\end{equation}
\end{theorem}

\begin{figure}[htbp]
\centering
\includegraphics[width=0.6\columnwidth]{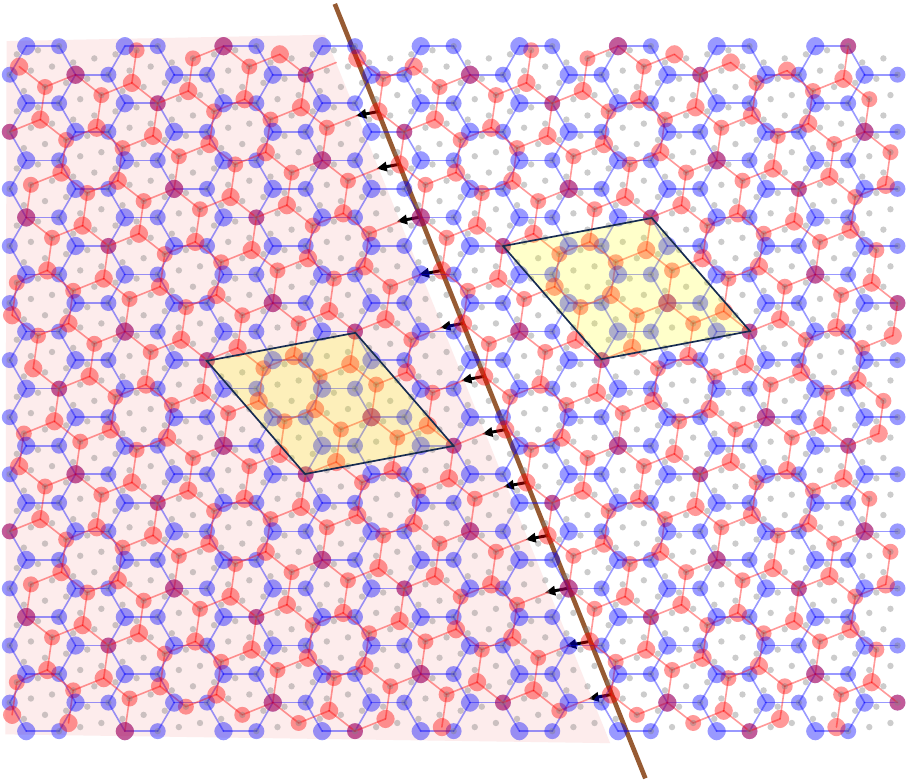}
\caption{Translational invariance of a $21.786789^\circ$ twisted BG. Translating the shaded region of the red lattice by a DSCL vector leaves the interface structure invariant and results in a shift in the CSL.}
\label{fig:dscl_shift}
\end{figure}
\fref{fig:dscl_shift} conveys the essence of \thref{thm:csl_shift} --- translating the shaded region of the red lattice by a DSCL vector (shown in arrows) leads to a CSL shift. However, since the interface structure is preserved, the interfacial energy is identical in the slipped and the non-slipped region, which leads us to the following definition --- \emph{an interface dislocation is a line of displacement discontinuity with Burgers vector equal to a DSCL vector.} Moreover, the DSCL translational invariance of the interface implies the GSFE is periodic with respect to the DSCL, and it suffices to describe the GSFE on a primitive unit cell of the DSCL.

\begin{figure}[htbp]
    \centering
    \subfloat[]
    {
        \includegraphics[width=0.48\textwidth]{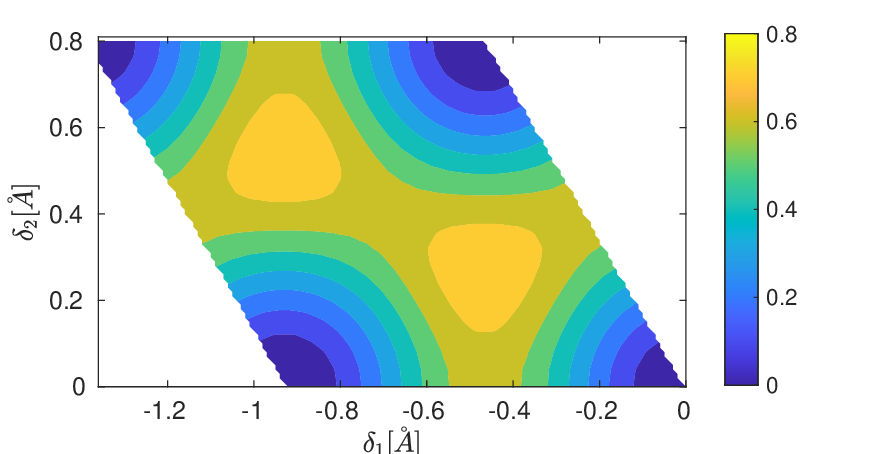}
        \label{fig:lammps_comp_gsfe}
    }
    \subfloat[]
    {
        \includegraphics[width=0.48\textwidth]{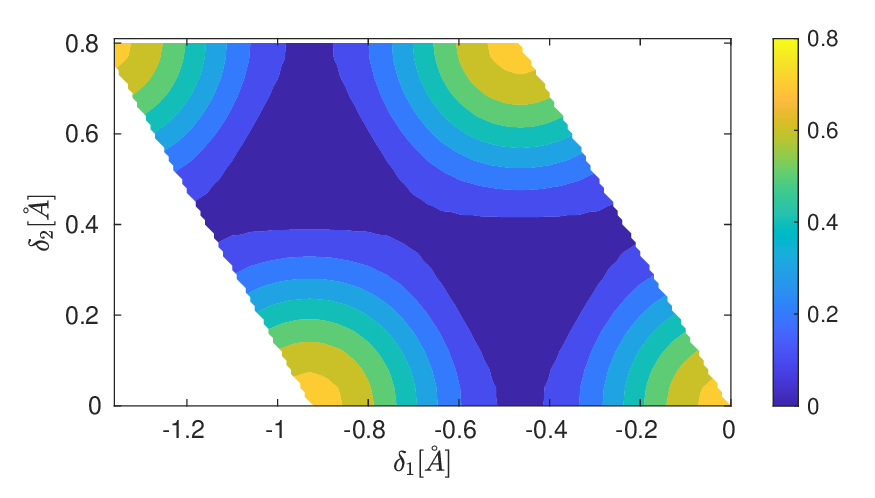}
        \label{fig:new_pot_21_78_gsfe}
    }
    \caption{GSFE [$\si{\milli\eV\per\angstrom\squared}$] plots of $\Sigma 7$ $\SI{21.786789}{\degree}$ twisted BG, computed in LAMMPS using \psubref{fig:lammps_comp_gsfe} the KC-1 parametrization at a $26\%$ out-of-plane compression, and \psubref{fig:new_pot_21_78_gsfe}  the KC-2 parametrization at $39\%$ out-of-plane compression.}
    \label{fig:gsfe}
    \end{figure}

Using the above definition, we can now revisit \fref{fig:schem} to reason the qualitative differences noted in \sref{sec:atomistics} between dislocations in the heterodeformed AB-stacked BG and the $\Sigma 7$ BG configurations. Since the DSCL of an AB-stacked BG is identical to the graphene lattice, its GSFE has the periodicity of graphene, and a dislocation in an AB-stacking is a lattice vector of graphene. However, full dislocations are not observed in an AB-stacking due to degenerate minima in the GSFE, i.e., the GSFE plotted on a primitive unit cell of the DSCL has more than one minimum. Instead, as noted in \sref{sec:atomistics_small}, partial dislocations form whose Burgers vector is the relative vector connecting two minima of the GSFE. Moving on to the $\Sigma 7$ BG, the magnitude of the smallest non-zero DSCL vector is equal to $2.46/\sqrt{7}=\SI{0.929792153}{\angstrom}$ (follows from \eqref{eqn:sigma} and \eqref{eqn:ABCD_relation}), which is indeed the displacement jump (see \fref{fig:twist_x_ux_old}) observed in the atomistic simulation with the KC-1 parametrization. This implies the line defects in $\Sigma 7$ BG, modeled using KC-1, are full interface dislocations. \fref{fig:lammps_comp_gsfe} shows the GSFE of the KC-1 modeled $\Sigma 7$ BG. The absence of degenerate minima in \fref{fig:lammps_comp_gsfe} justifies the absence of partials in \fref{fig:twist_x_ux_old}. However, the GSFE of the KC-2 modeled $\Sigma 7$ BG, shown in \fref{fig:new_pot_21_78_gsfe}, features two degenerate minima, which suggests the interface dislocations in \fref{fig:twist_x_ux_new} are partials. The magnitude of the Burgers vector of the partials can be inferred from \fref{fig:new_pot_21_78_gsfe} as $2.46/(\sqrt{3}\sqrt{7})=\SI{0.536816}{\angstrom}$. In addition to the Burgers vectors, the GSFE also determines the arrangement of the dislocation network. Recall from \sref{sec:atomistics_large}, that the KC-1 parametrization resulted in a honeycomb dislocation network that separates defect-free $\Sigma 7$-stacked hexagonal regions. We assert that \emph{the number of sides of the low-energy region is determined by the number of nearest neighbor GSFE minimizers of a minimizer}. This assertion is corroborated by GSFE plots in \fref{fig:gsfe} --- a GSFE minimizer in the KC-2 parameterization has three nearest neighbor minimizers, and therefore, the resulting dislocation network is formed by triangular defect-free $\Sigma 7$-stackings.

\begin{figure}[t]
    \centering
        \includegraphics[width=0.48\textwidth]{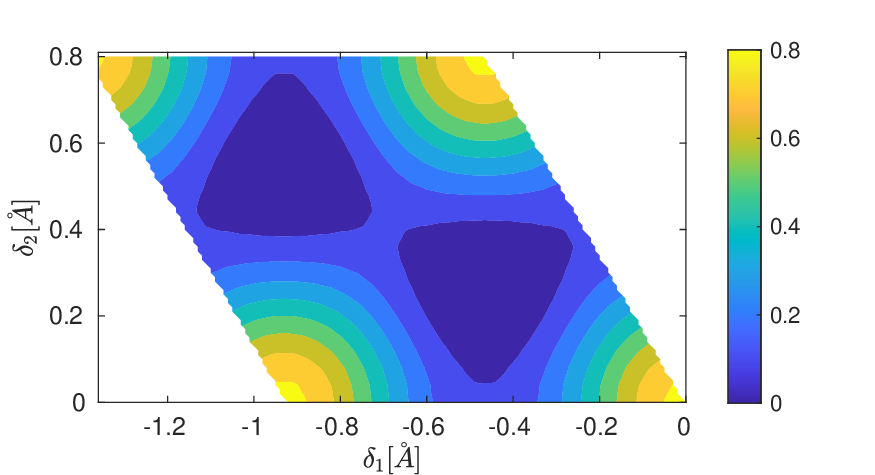}
    \caption{Density functional theory-generated GSFE [$\si{\milli\eV\per\angstrom\squared}$] of a $\Sigma 7$ $\SI{21.786789}{\degree}$ twisted BG at a $26\%$ out-of-plane compression.}
    \label{fig:DFT_gsfe}
    \end{figure}

Recall from \sref{sec:atomistics_large} that the step character of the displacement ``jump'' is accentuated as the out-of-plane compression increases. This is because as the two layers are compressed the difference between the maximum and the minimum values of the GSFE, which we will refer to as the \emph{GSFE range width}, increases. We noticed that compared to the KC-1 modeled BG, the KC-2 modeled $\Sigma 7$ BG has a smaller GSFE range width. This was the primary reason we chose a larger out-of-plane compression ($39\%$) in the simulation using the KC-2 model so that the GSFE range widths of the two parametrizations are equal to $\SI{0.8}{\milli\eV \per \angstrom\squared}$, as shown in \fref{fig:gsfe}. Although the GSFE range widths match, the interface dislocations of the KC-2 model are more diffused as the Burgers vector of the partial is smaller by a factor of $\sqrt{3}$. Finally, we verify using density functional theory (DFT) calculations that the KC-2 is superior to KC-1. The DFT-generated GSFE (see \ref{sec:dft}), plotted in \fref{fig:DFT_gsfe}, compares qualitatively well with the GSFE in \fref{fig:new_pot_21_78_gsfe}. However, from a modeling perspective, the two models are equally valuable as they demonstrate the links between bicrystallography, GSFE, and the properties of interface dislocations.

\section{A generalized Frenkel--Kontorova (GFK) model for 2D heterointerfaces}
\label{sec:continuum}
The goal of this section is to build a continuum model to predict structural relaxation in heterostructures subjected to arbitrary heterodeformations by generalizing the Frenkel--Kontorova model of \citet{Koshino_2017}. In \sref{sec:numerics}, we apply the GFK model to predict structural relaxation in heterodeformed BGs. The kinematics of the GFK model is inspired by the framework of large deformation crystal plasticity \citep{clayton2010,ADMAL2018,Junyan_2021,Himanshu_2022}, wherein dislocations are defined with respect to a defect-free natural configuration. 

\subsection{Kinematics}
\label{sec:kinematics}
Consider an interface formed by two 2D (multi) lattices. For simplicity, we will ignore out-of-plane displacement and assume the lattices occupy regions $\Omega^{\rm ref}_{\rm t}$ and $\Omega^{\rm ref}_{\rm b}$ in the 2D Euclidean point space $\mathbb R^2$.\footnote{The out-of-plane displacement during atomic reconstruction is an important feature recently studied by \citet{srolovitz_2016}. While incorporating the out-of-plane displacement into our continuum model is conceptually straightforward, we chose to ignore it to better convey the GFK model.} The variable $\alpha$ is used to index the top ($\rm t$) and bottom ($\rm b$) layers, i.e., $\alpha=\rm t$ or $\rm b$. The role of a reference configuration is to measure displacements relative to a fixed configuration, and its choice should not affect the predictions of the model. Since our goal is to isolate and predict displacements associated with atomic reconstruction --- as opposed to large-scale deformation --- the reference configurations are chosen such that the lattices in $\Omega^{\rm ref}_\alpha$ are twisted relative to each other or are independently strained uniformly such that they are marginally misaligned relative to a low energy moir\'e configuration.\footnote{This choice is also motivated by the 'tear and stack' technique \citep{kim2016van} to control the twist in a BG.} From \sref{sec:atomistics}, we know that such reference configurations are not stable and undergo atomic reconstruction by nucleating interface dislocations. The goal of this section is to construct \emph{frame-invariant} kinematic measures that quantify the elastic and vdW energies responsible for the atomic reconstruction. 

Let $\bm \phi_\alpha:\Omega^{\rm ref}_\alpha \times [0,\infty] \to \mathbb R^2$ ($\alpha=\rm t,\rm b$) denote time-dependent deformation maps associated with the atomic reconstructions of the respective lattices, measured relative to their reference configurations. The deformed configuration to which $\bm \phi_\alpha$ maps to will be denoted by $\Omega_\alpha$. Adopting the convention of continuum mechanics, we use $\bm X_\alpha$ to denote an arbitrary material point in $\Omega_\alpha^{\rm ref}$. The gradients of the deformation maps are given by $\bm F_\alpha:=\nabla \bm \phi_\alpha$. At this stage, it is useful to connect to the heterostrained moir\'e example of \sref{sec:atomistics}, wherein a $\Sigma 7$ moir\'e twisted BG ($21.786789^\circ$ twist relative to the AB stacking) when subjected to principal stretches of $1.05\%$ and $-0.2\%$, was observed to undergo atomic reconstruction under PBCs. For a continuum analog of this system, the reference configuration will reflect the atomistic system prior to the energy minimization, and $\bm \phi_\alpha-\bm X_\alpha$ corresponds to the displacements due to energy minimization. If the two lattices in the reference configurations are allowed to relax in the absence of external loads, they will a) return to their respective planar strain-free configurations, and b)twist by an angle that minimizes the vdW energy (as a function of the misorientation angle). In plasticity, this relaxed configuration is commonly referred to as a \emph{natural configuration}. 

\begin{figure}[t]
\centering
\includegraphics[width=\columnwidth]{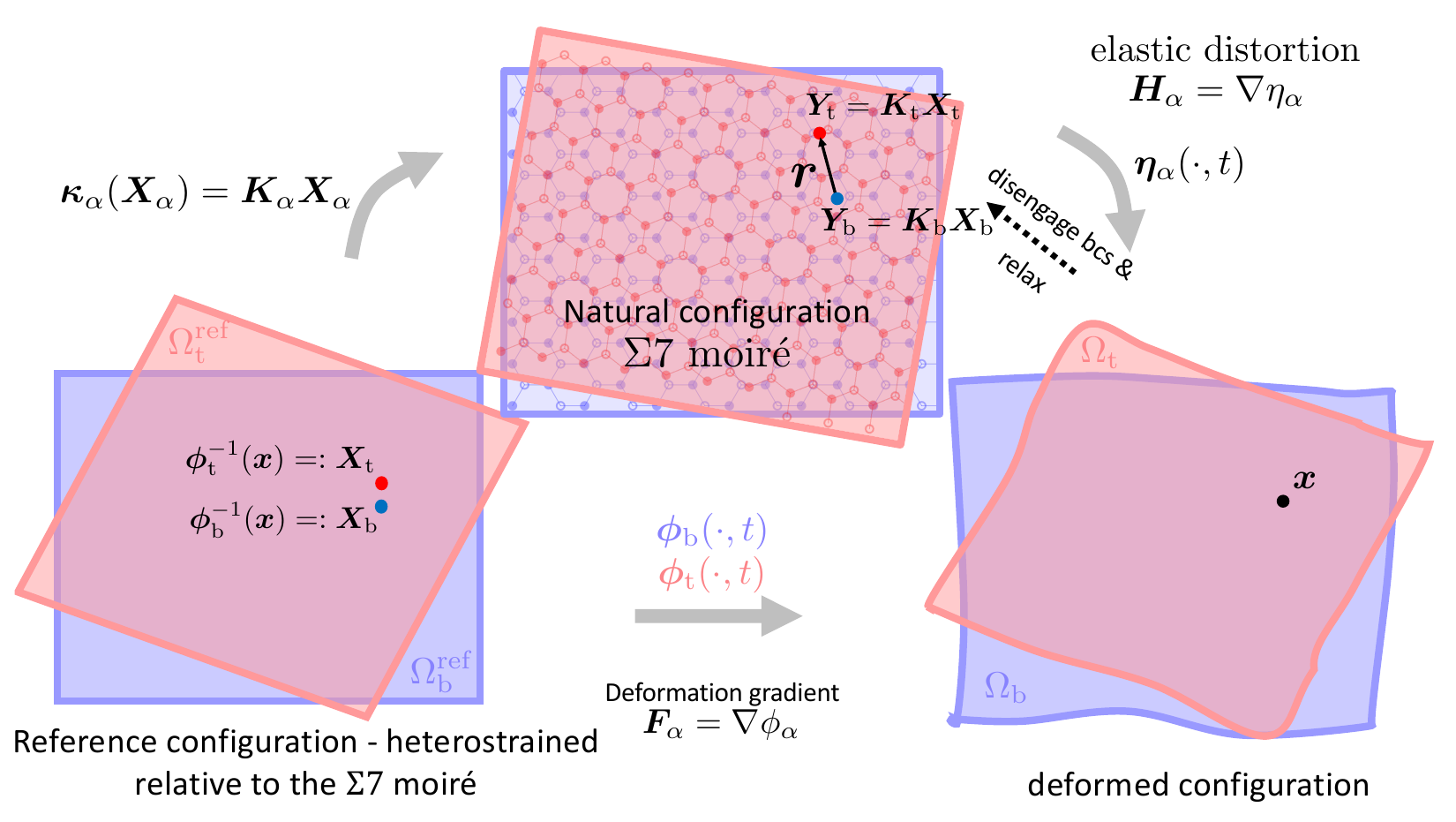}
\caption{A schematic of the reference (left), natural (middle), and deformed (right) configurations of the GFK model.}
\label{fig:cont_geom}
\end{figure}
The idea of a natural configuration plays a central role in our framework as we will show that deformation measures defined with respect to the natural configuration are frame-invariant and independent of the choice of the reference configuration. Employing the language of crystal plasticity theories, we let $\bm K_\alpha$ represent the map from the tangent space of $\Omega_\alpha^{\rm ref}$ to that of the natural configuration. In this work, $\bm K_\alpha$ is a constant tensor, and $\bm K_\alpha^{-1}$ can be interpreted as the average deformation gradient (relative to the natural configuration) an experimentalist imposes. The mapping from the reference configuration to the natural configuration is given by $\bm \kappa_{\alpha}(\bm X_\alpha):=\bm K_\alpha \bm X_\alpha$. $\Omega_\alpha^{\rm n}$ is used to denote a lattice in the natural configuration and points in $\Omega_\alpha^{\rm n}$ will be denoted by $\bm Y_\alpha$. Furthermore, we use $\bm \eta_\alpha$ to denote the mapping from the natural configuration to the deformed configuration. By construction, we have
\begin{equation}
    \bm \phi_\alpha = \bm \eta_\alpha \circ \bm \kappa_\alpha,
    \label{eqn:composition}
\end{equation}
where $\circ$ denotes function composition. From \eqref{eqn:composition}, the following relationship between the gradients of the deformation maps follow:
\begin{equation}
   \bm F_\alpha = \bm H_\alpha \bm K_\alpha, \text{ where } \bm H_\alpha:=\nabla \bm \eta_\alpha.
   \label{eqn:FeFp}
\end{equation}
Note that the gradient in \eqref{eqn:FeFp} is with respect to $\bm Y_\alpha$.
Unlike $\bm K_\alpha$,  $\bm H_\alpha$ is a time-dependent field, and its inverse describes the relaxation of a local neighborhood of a point $\bm x \in \Omega_\alpha$ in the absence of external loads. In the context of our heterostrained moir\'e example, the natural configuration is the $\Sigma 7$ moir\'e since the two lattices are strain-free and the interfacial energy is minimum in neighborhoods of small hetero-strains and twists. Moreover, $\Kb \equiv \bm I$ and $\Kt^{-1}$ is equal to the biaxial strain, given in \eqref{eqn:U_largeTwist}.

We will now construct frame-invariant kinematic measures to quantify the elastic and vdW energies. 
Since the elastic energy due to atomic reconstruction originates from the strains in the lattices measured with respect to a strain-free natural configuration, the relevant frame-invariant kinematic measure is the Cauchy--Green deformation tensor $\bm C_\alpha:= \bm H_\alpha^{\rm T} \bm H_\alpha$. From \eqref{eqn:FeFp}, $\bm C_\alpha$ can be written as
\begin{equation}
    \bm C_\alpha = \bm K_\alpha^{-\rm T} \bm F^{\rm T}\bm F \bm K_\alpha^{-1}.
\end{equation}
On the other hand, the vdW energy originates from the interaction between lattices in the region $\Omega_{\rm t} \cap \Omega_{\rm b}$. The vdW energy is described by the relative translation between the two lattices when allowed to relax to the natural configuration. Therefore, the vdW energy density at a point $\bm x \in \Omega_{\rm t} \cap \Omega_{\rm b}$ will be expressed as a function of the relative vector
\begin{equation} 
\bm r(\bm x,t)= \Kt \Xt-\Kb \Xb, \text{ where }
    \bm X_\alpha:= \bm \phi_\alpha^{-1}(\bm x,t).
    \label{eqn:r}
\end{equation}
Summarizing, we have two frame-invariant kinematic measures, $\bm C_\alpha$ and $\bm r$, expressed in terms of the deformation map $\phi_\alpha$ for given $\bm K_\alpha$, that characterize elastic and vdW energies, respectively. In the next section, we will describe the constitutive laws for the elastic and vdW energies in terms of $\bm C_\alpha$ and $\bm r$.

\subsection{Constitutive law}
\label{sec:constitutive}
In this section, we construct a frame-invariant energy functional for the GFK model. For  prescribed average heterodeformations ($\bm K_\alpha^{-1}$), the total energy functional $\mathcal E$ is additively decomposed as 
\begin{equation}
\mathcal E[\phit,\phib]= \mathcal E_{\rm el}+\mathcal E_{\rm vdW},
\label{eqn:totalEnergy}
\end{equation}
into elastic and interfacial energies. Since the elastic energy corresponds to elastic distortions relative to the natural configurations, we assume $\mathcal E_{\rm el}$ to be an integral of an elastic energy density (per unit area in the natural configuration) $e_{\rm el}$ over the natural configuration:
\begin{equation}
    \mathcal E_{\rm el}[\phit, \phib] = \sum_{\alpha=\mathrm t, \mathrm b}
    \int_{\Omega_\alpha^{\rm n}} 
    e_{\rm el}(\bm E_\alpha;\alpha) \, d\bm Y_\alpha, \quad  \quad \text{where } \bm E_{\alpha}=(\bm C_\alpha-\bm I)/2
    \label{eqn:elasticEnergy}
\end{equation}
is the frame-invariant Lagrangian strain tensor, and $e_{\rm el}(\bullet; \alpha)$ is the elastic energy density of the $\alpha$-th layer. For example, a Saint Venant--Kirchhoff elastic energy density is of of the form $e_{\rm el}=\mathbb C_\alpha \bm E_\alpha \cdot \bm E_\alpha/2$, where $\mathbb C_\alpha$ is the fourth-order elasticity tensor of the $\alpha$-lattice.

\begin{figure}[t]
    \centering
    \subfloat[]
    {
        \includegraphics[width=0.45\textwidth]{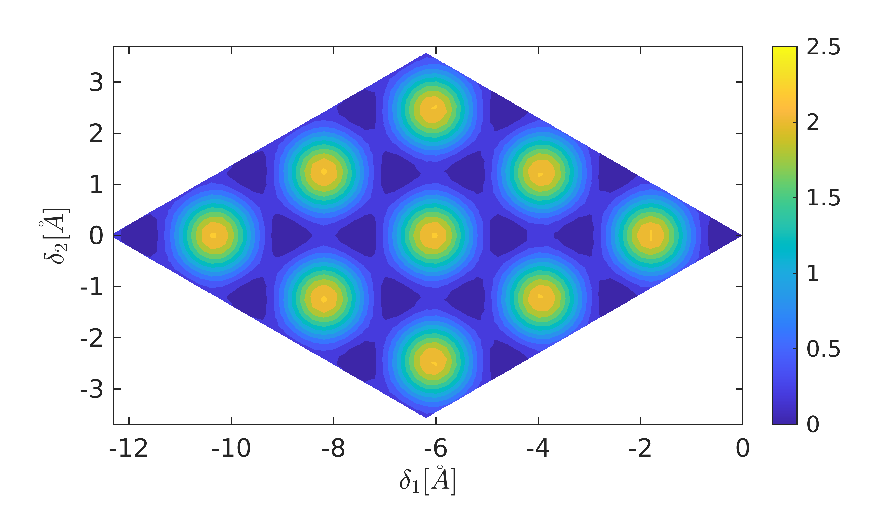}
        \label{fig:GSFE_global_en_den_con}
    }
    \subfloat[]
    {
        \includegraphics[width=0.45\textwidth]{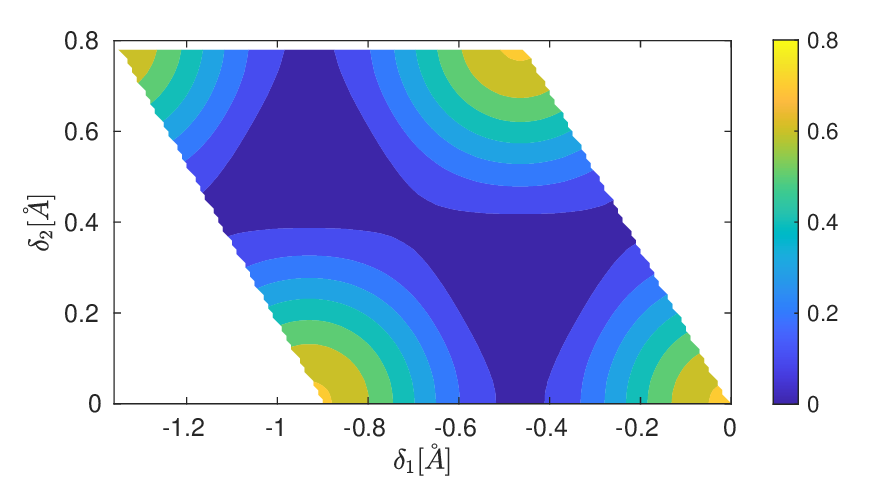}
        \label{fig:GSFE_local_en_den_con}
    }
    \caption{Plots of $e_{\rm vdW}$ for \psubref{fig:GSFE_global_en_den_con} AB-stacked $\Sigma 1$ and  \psubref{fig:GSFE_local_en_den_con} $\Sigma 7$ configurations, computed using \eqref{eq:inter_en} and plotted such that the minimum is zero.} 
    \label{fig:GSFE_cont}
    \end{figure}

The interaction energy term $\mathcal E_{\rm vdW}$ measures the changes in the vdW energy due to relative translations between the lattices in the natural configuration. Since the lattices interact in the overlapping region $\Omega_{\rm t} \cap \Omega_{\rm b}$ of the deformed configuration, we express $\mathcal E_{\rm vdW}$ as an integral over $\Omega_{\rm t} \cap \Omega_{\rm b}$ of a vdW energy density $e_{\rm vdW}$ --- measured per unit area in the natural configuration. From an atomistic viewpoint, $e_{\rm vdW}$ is the GSFE density introduced in \sref{sec:atomistics}, and is expressed as a function of the frame-invariant relative vector $\bm r$ introduced in \eqref{eqn:r}. Therefore, 
\begin{equation}
\mathcal E_{\rm vdW}[\phit, \phib] =  
\frac{1}{2}
\sum_{\alpha=\rm t, \rm b}
\int_{\Omega_{\rm t} \cap \Omega_{\rm b}} 
(\det \bm H_\alpha)^{-1} e_{\rm vdW}(\bm r(\bm x_\alpha)) \, d\bm x_\alpha.
\label{eqn:vdWEnergy}
\end{equation}
Note that the factor $(\det \bm H_\alpha)^{-1}$ is necessary because the integration is over the deformed configuration as opposed to the natural configuration. As a result, the interaction energy has to be split evenly between the two lattices leading to the factor of $1/2$. From \sref{sec:atomistics}, we know that $e_{\rm vdW}$ has the periodicity of the DSCL corresponding to the natural configuration. For example, vdW energy densities for the AB stacked-$\Sigma 1$ and the $\Sigma 7$ natural configurations have the form 
\begin{equation}
    e_{\rm vdW}(\bm r) = \pm 2 \bm v_0\sum_{p=1}^{3}\cos \left (2 \pi 
        \bm{\mathcal d}^p \cdot (\bm r+\bm t)
    \right) + c,
    \label{eq:inter_en}
\end{equation}
where $\bm{\mathcal d}^1$ and $\bm{\mathcal d}^2$ are basis vectors of the reciprocal lattice, $\bm{\mathcal d}^3=-\left(\bm{\mathcal d}^1+\bm{\mathcal d}^2\right)$, $\bm t$ is a translation vector, and $c$ and $v_0$ are constants. By comparing \eqref{eq:inter_en} to the atomistics/first principles GSFE densities plotted in \frefs{fig:AB_gsfe}{fig:DFT_gsfe} we obtain $v_0$ and $\bm t$ corresponding to the $\Sigma 1$ and $\Sigma 7$ configurations. Plots of the resulting $e_{\rm vdW}$, shown in \fref{fig:GSFE_cont}, compare well with those in \fref{fig:DFT_gsfe}.

Next, we propose an evolution law for the unknown fields, $\phit$  and $\phib$, as a gradient flow of $\mathcal E$:
\begin{equation}
    m \dot{\bm \phi}_\alpha = -\updelta_{\bm \phi_\alpha} \mathcal E, \quad \alpha = \rm t, \rm b
    \label{eqn:gradientFlow}
\end{equation}
where $m_\alpha$ is a prescribed mobility, and $\updelta_{\bm \phi_\alpha}$ denotes variation with respect to $\bm \phi_\alpha$. 

\subsection{Derivation of the governing equations of the GFK model}
In this section, we derive the governing equations of the GFK model by calculating the variational derivative in \eqref{eqn:gradientFlow}. The derivation is applicable to finite systems, a notable departure from earlier works, which focused on infinite systems modeled using PBCs. In addition to the critical role vdW energy plays in the structural relaxation of 2D heterostructures, we will show that it manifests as surface tension, which contributes towards configurational forces on the two lattices.

To compute the variation with respect to $\bm \phi_\alpha$, we transform the integrals in \eqref{eqn:elasticEnergy} and \eqref{eqn:vdWEnergy} to the reference configurations. We begin by rewriting the elastic energy in \eqref{eqn:elasticEnergy} by noting that $d\bm Y_\alpha=(\det \bm K_\alpha) d\bm X_\alpha$:
\begin{equation}
    \mathcal E_{\rm el}[\phit, \phib] = \sum_{\alpha=\mathrm t, \mathrm b}
    \int_{\Omega_\alpha^{\rm ref}} 
    e_{\rm el}(\bm E_\alpha;\alpha) J_\alpha \, d\bm X_\alpha,
    \label{eqn:elasticEnergyReference}
\end{equation}
where $J_\alpha:= \det \bm K_\alpha$. Taking the variation of $\mathcal E_{\rm el}$ in \eqref{eqn:elasticEnergyReference} with respect to $\bm \phi_\alpha$, we obtain 
\begin{align}
    -\updelta_{\bm \phi_\alpha} \mathcal E_{\rm el} = J_\alpha \divr(\bm P_\alpha) \text{ in $\Omega_\alpha^{\rm ref}$, where }
    \bm P_\alpha:= \bm H_\alpha \bm \nabla \bm e_{\rm el} \bm K_\alpha^{-\rm T}.
    \label{eqn:variation_elastic}
\end{align}
The tensor $\bm P_\alpha$ is the 2D analog of the elastic Piola--Kirchhoff stress, which measures force in $\Omega_\alpha$ measured per unit length in $\Omega_\alpha^{\rm ref}$. In addition, the variational derivative also yields the usual expression for the traction on the boundary $\Gamma_\alpha$ of $\Omega_\alpha$ as $\bm P_\alpha \bm n_\alpha$, where $\bm n_\alpha$ is a outward unit vector normal to $\Gamma_\alpha^{\rm ref}$.

\begin{figure}[t]
\centering
\includegraphics[width=0.6\columnwidth]{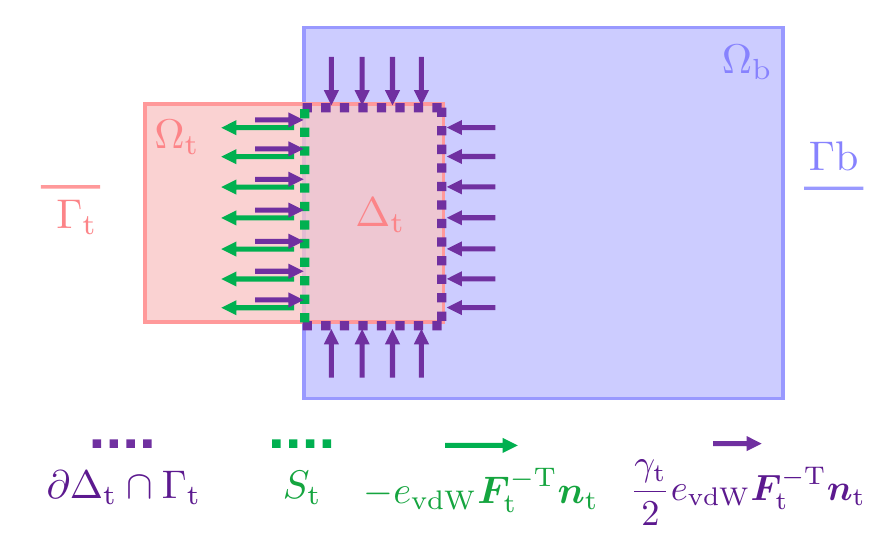}
\caption{A schematic showing two types of tractions (purple and green arrows) on the boundary of the overlapping region, $\Delta_{\rm t} \subset \Omega_{\rm t}$, due to surface tension associated with vdW interactions in a finite 2D heterostructure. The traction forces balance the thermodynamic driving forces that act to increase the region of overlap. The purple arrows are compressive forces associated with the overlapping region's tendency to dilate to increase its area. The green arrows are configurational forces on $S_{\rm t}$ that balance the thermodynamic forces conjugate to aerial changes in $\Delta_{\rm t}$ due to the translation of $\Omega_{\rm t}$ into $\Omega_{\rm b}$.}
\label{fig:forces}
\end{figure}
Compared to \eqref{eqn:variation_elastic}, calculating the variation of $\mathcal E_{\rm vdW}$ in \eqref{eqn:vdWEnergy} is a delicate exercise due to a) the presence of the inverse function $\bm \phi_\alpha^{-1}$, and b) the domain of integration in \eqref{eqn:vdWEnergy} is a part of the deformed configuration and is therefore sensitive to $\bm \phi_\alpha$. We begin with $\updelta_{\phit} \mathcal E_{\rm vdW}$. To eliminate the dependence on $\phit^{-1}$, we transform the two domains of integration in \eqref{eqn:vdWEnergy} to $\Delta^{\rm ref}_{\rm t}:=\phit^{-1}(\Omega_{\rm t} \cap \Omega_{\rm b})$ in the reference configuration by noting that $d\xt=d\xb=(\det \Ft) d\Xt$, resulting in
\begin{equation}
    \mathcal E_{\rm vdW}[\phit, \phib] 
    =\frac{J_{\rm t}}{2}
    \int_{\Delta^{\rm ref}_{\rm t}} 
    \left (
    1+\underbrace{\det (\Ht\Hb^{-1})}_{\gamma_{\rm t}}
    \right)
    e_{\rm vdW}(\rt(\Xt)) \, d\Xt.
    \label{eqn:transform_vdW}
\end{equation}
where $\bm r_{\alpha}(\bm X_\alpha):=\bm r(\bm x)|_{\bm x=\bm \phi_{\alpha}(\bm X_\alpha)}$, i.e. 
\begin{equation}
    \rt(\Xt) = \Kt \Xt - \Kb (\phib^{-1} \circ \phit(\Xt)) \text{ and }
    \rb(\Xb) = \Kt (\phit^{-1} \circ \phib(\Xb))-\Kb \Xb .
    \label{eqn:r_alpha}
\end{equation}
From \eqref{eqn:r_alpha}, we note that the integrand in \eqref{eqn:transform_vdW} depends on $\phit$ and its gradient but not its inverse, as desired. The variational derivative of \eqref{eqn:transform_vdW} with respect to $\phit$ now follows:
\begin{equation}
-\updelta_{\phit} \mathcal E_{\rm vdW} = J_{\rm t}\left[ \frac{1+\gamma_{\rm t}}{2} \Hb^{-\rm T} \nabla e_{\rm vdW} + \divr
\left (\frac{\gamma_{\rm t}}{2} e_{\rm vdW} \Ft^{-\rm T}
\right )
\right] \text{ in $\Delta_{\rm t}^{\rm ref}$.}
\label{eqn:variation_vdWt}
\end{equation}
along with two kinds of traction forces
\begin{subequations}
\begin{align}
    \frac{\gamma_{\rm t}}{2} e_{\rm vdW} \Ft^{-\rm T} \bm n_{\rm t}  &\text{ on } \partial \Delta^{\rm ref}_{\rm t},
    \label{eqn:traction1_vdW}\\
    -\frac{1+\gamma_{\rm t}}{2} e_{\rm vdW} \Ft^{-\rm T} \bm n_{\rm t}  &\text{ on } S_{\rm t} := \partial \Delta_{\rm t}^{\rm ref} - \Gamma_{\rm t}^{\rm ref}.
    \label{eqn:traction2_vdW}
\end{align}
\label{eqn:traction_vdW}
\end{subequations}

The expression $\nabla e_{\rm vdW}$ in \eqref{eqn:variation_vdWt} seeks to increase areas of high commensurability, and is responsible for the formation of interface dislocations. The term $e_{\rm vdW} \gamma_{\rm t}/2$ in \eqref{eqn:variation_vdWt} is the surface tension/pressure that is conjugate to aerial changes in $\Omega_{\rm t}$, and \eqref{eqn:traction1_vdW} is the corresponding traction. The traction in \eqref{eqn:traction1_vdW} acts to counter the thermodynamic driving force that tends to dilate $\Omega_{\rm t}$ --- since $e_{\rm vdW}$ is negative --- to maximize the area of overlap with $\Omega_{\rm b}$. Therefore, the traction in $\eqref{eqn:traction1_vdW}$ is compressive, as the purple arrows in \fref{fig:forces} depict. On the other hand, the traction in \eqref{eqn:traction2_vdW}, shown in green in \fref{fig:forces}, is a configurational force that acts on the part of $\partial \Delta_{\rm t}^{\rm ref}$ that belongs to the interior of $\Omega_{\rm t}$. It works to oppose the thermodynamic driving force that translates $\Omega_{\rm t}$ into $\Omega_{\rm b}$, thereby increasing the area of overlap.\footnote{Mathematically, the configurational force on $S_{\rm t}$ arises due to the dependence of 
$\Delta_{\rm t}^{\rm ref}$ (the domain of integration in \eqref{eqn:transform_vdW}) on $\phit$.} Notice that the boundary $S_{\rm t}$ experiences both tractions mentioned in \eqref{eqn:traction_vdW}, as shown in \fref{fig:forces}. Therefore, the total traction on $\partial \Delta_{\rm t}$ due to vdW interactions is given by
\begin{subequations}
    \begin{align}
        -e_{\rm vdW} \Ft^{-\rm T} \bm n_{\rm t}  &\text{ on } S_{\rm t}, \text{ and } \label{eqn:totalTraction1_vdW}\\
        \frac{\gamma_{\rm t}}{2} e_{\rm vdW} \Ft^{-\rm T} \bm n_{\rm t}  &\text{ on } \partial \Delta^{\rm ref}_{\rm t} -S_{\rm t}. \label{eqn:totalTraction2_vdW}
    \end{align}
    \label{eqn:totalTraction_vdW}
\end{subequations}

Substituting \eqref{eqn:variation_elastic} and \eqref{eqn:variation_vdWt} into \eqref{eqn:gradientFlow}, results in a governing equation for $\Omega_{\rm t}$:
\begin{equation}
    m_{\rm t} \dot{\bm \phi}_{\rm t} = 
    \begin{cases}
    \divrt \left (\bm P_{\rm t} + 
    \frac{\gamma_{\rm t}}{2} e_{\rm vdW} \Ft^{-\rm T}
\right ) + \frac{1+\gamma_{\rm t}}{2}\Hb^{-\rm T} \nabla e_{\rm vdW}(\bm r_{\rm t}) \text{ in $\Delta_{\rm t}$}, \\
    \divrt(\bm P_{\rm t})\text{ in $\Omega_{\rm t}-\Delta_{\rm t}$},
    \end{cases}
    \label{eqn:fullEquation_1}
\end{equation}
where $m_\alpha:= J_{\alpha}^{-1} m$, along with traction boundary conditions in \eqref{eqn:totalTraction_vdW}. Similarly, the governing equation for $\Omega_{\rm b}$ is given by
\begin{equation}
    m_{\rm b} \dot{\bm \phi}_{\rm b} = 
    \begin{cases}
    \divrb
    \left (\bm P_{\rm b} +
    \frac{\gamma_{\rm b}}{2} e_{\rm vdW} \Fb^{-\rm T}
    \right)-\frac{1+\gamma_{\rm b}}{2}\Ht^{-\rm T} \nabla e_{\rm vdW}(\rb) \text{ in $\Delta_{\rm b}$},\\
    \divrb(\bm P_{\rm b}) \text{ in $\Omega_{\rm b}-\Delta_{\rm b}$},
    \end{cases}
    \label{eqn:fullEquation_2}
\end{equation}
where $\gamma_{\rm b}:=\det (\Hb\Ht^{-1})$, along with corresponding traction boundary conditions. A notable feature of the governing equations is that the total stress now includes a contribution from surface tension, which was absent in previous Frenkel--Kontorova models that were developed for infinite systems. While the role of surface tension may be ignored for infinite systems, we expect it to play an important role in finite systems, wherein sliding between the constituent 2D lattices is enhanced. It is worth pointing out that the two key features of our model --- surface tension and frame-invariance\footnote{Indeed, under an arbitrary superposed rigid body displacement given by a constant rotation tensor $\bm R$ and a constant vector $\bm c$, the fields transform to 
\begin{equation}
    \tilde{\bm \phi}_\alpha=\bm R \bm \phi_\alpha + \bm c; \quad
    \tilde{\bm F}_\alpha = \bm R \bm F_\alpha (\implies \tilde{\bm H}_\alpha = \bm R \bm H_\alpha); \quad
    \tilde{\bm \sigma}_\alpha = \bm \sigma_\alpha; \quad
    \tilde{\bm K}_\alpha = \bm K_\alpha; \quad
    \tilde{\bm r}_\alpha = \bm r_\alpha,
\end{equation}
and continue to satisfy \eqref{eq:final_eqn}.\label{fn:srb}
} --- are a consequence of the model's geometrically nonlinear kinematic framework.

Next, we will focus our attention on using the GFK model to simulate atomic reconstruction in infinite 2D heterostructures and compare its predictions with atomistic simulation results of \sref{sec:atomistics}. To this end, we simplify our model for numerical implementation by ignoring the surface tension terms and assuming $J_\alpha \approx 1$, resulting in
\begin{subequations}
\begin{align}
    m \dot{\bm \phi}_{\rm t} &= \divrt(\bm P_{\rm t}) + \Hb^{-\rm T} \nabla e_{\rm vdW}(\bm r_{\rm t}), \\
    m \dot{\bm \phi}_{\rm b} &= \divrb(\bm P_{\rm b}) -  \Ht^{-\rm T} \nabla e_{\rm vdW}(\bm r_{\rm b}).
\end{align}
\label{eq:final_eqn}
\end{subequations}
Using \eqref{eqn:r_alpha} and recalling from \eqref{eqn:FeFp} that $\bm H_\alpha= \bm F_\alpha \bm K_\alpha^{-1}$, the right-hand-side of \eqref{eq:final_eqn} can be expressed entirely in terms of the unknown $\bm \phi_{\alpha}$, its gradient, and its inverse. However, the dependence on the inverse is impractical for numerical computation. Therefore, we resort to approximating\footnote{
$\rt$ and $\rb $ can be expressed as
\begin{subequations}
\begin{align}
    \rt &= \Kt\Xt-\Kb\Xb  =(\Kt-\Kb) \Xt + \Kb (\Xt-\Xb),\label{eqn:rt}\\
    \rb &= \Kt \Xt-\Kb \Xb 
    =\Kt (\Xt-\Xb) + (\Kt -\Kb) \Xb.\label{eqn:rb}
\end{align}
\end{subequations}
Since $\phit(\Xt)=\phib(\Xb)$, it follows that
\begin{align*}
    \phit(\Xt)-\phib(\Xt) &= \phib(\Xb)-\phib(\Xt) \approx \Fb (\Xb-\Xt),\\
    \phit(\Xb)-\phib(\Xb) &= \phit(\Xb)-\phit(\Xt) \approx \Ft (\Xb-\Xt),
\end{align*}
which imply 
\begin{subequations}
\begin{align}
    \Xt-\Xb &\approx -\Fb^{-1}(\phit(\Xt)-\phib(\Xt)),\label{eqn:approx_t}\\
    \Xt-\Xb &\approx -\Ft^{-1}(\phit(\Xb)-\phib(\Xb)). \label{eqn:approx_b}
\end{align}
\label{eqn:approx}
\end{subequations}
Equation \eqref{eqn:rt} and the approximation \eqref{eqn:approx_t} yield \eqref{eqn:rt_approx}. Similarly, \eqref{eqn:rb} and \eqref{eqn:approx_b} result in \eqref{eqn:rb_approx}.
} $\rt$ and $\rb$ as 
\begin{subequations}
\begin{align}
    \rt &\approx (\Kt -\Kb) \Xt -  \Hb^{-1} (\phit(\Xt)-\phib(\Xt)),\label{eqn:rt_approx}\\
    \rb &\approx (\Kt -\Kb) \Xb -  \Ht^{-1} (\phit(\Xb)-\phib(\Xb))). \label{eqn:rb_approx}
\end{align}
\label{eqn:r_approx}
\end{subequations}
The above approximation preserves the frame-invariance of the model. In the next section, we will present a numerical implementation of \eqref{eq:final_eqn} with the $\bm r_\alpha$-approximation in \eqref{eqn:r_approx}.

\section{Numerical implementation of the GFK model and comparison with atomistics}
\label{sec:numerics}
The goal of this section is to implement the GFK model by solving \eqref{eq:final_eqn} numerically under PBCs, simulate atomic reconstruction in heterostrained small- and large-twist BGs, and validate the continuum model by comparing to atomistics. While the GFK model applies to finite and infinite domains, we restrict the numerical implementation to periodic systems as the primary goal is to compare to the periodic atomistic simulations of \sref{sec:atomistics}.

Inputs to the model include --- 
\begin{itemize}
    \item the top lattice $\mathcal A$ and the bottom lattice $\mathcal B$ of the natural configuration, which form a moir\'e supercell. For example, the AB-stacking and the $\Sigma 7$ configurations; 
    \item the average heterodeformations $\bm K_\alpha^{-1}$ imposed on the natural configuration, which determine the reference configuration. We take $\bm K_{\rm b}=\bm I$, and $\bm K_{\rm t}^{-1}$ is chosen from the set of heterodeformations that ensure the reference configuration satisfies PBCs, i.e. lattices $\mathcal A$  and $\bm K_{\rm b}^{-1}\mathcal B$ share a CSL. A primitive unit cell of this CSL is chosen as the reference configuration and the simulation domain. For example, when AB-stacking is the natural configuration, $\bm K_{\rm b}^{-1}$ is a $0.2992634^\circ$ twist or the heterostrain in \eqref{eqn:U_smallTwist}. When $\Sigma 7$ is the natural configuration, $\bm K_{\rm b}^{-1}$ is a $0.170076^\circ$ twist or the strain in \eqref{eqn:U_largeTwist}.
    \item isotropic elastic constants of graphene: $\lambda=\SI{3.5}{\eV\per\angstrom\squared}$ and $\mu=\SI{7.8}{\eV\per\angstrom\squared}$; and 
    \item the constant $v_0$ of the vdW energy density in \eqref{eq:inter_en}. For the AB-stacking, $v_0$ is equal to  $\SI{0.25}{\milli\eV\per\angstrom\squared}$, and for the $\Sigma 7$ configuration, it is  set to $\SI{-0.08}{\milli\eV\per\angstrom\squared}$ and $\SI{0.08}{\milli\eV\per\angstrom\squared}$ corresponding to the  KC-1 and KC-2 potentials, respectively.
    \item the mobilities $m_\alpha$, which are chosen as unity.
\end{itemize}

\subsection{Numerical method} 
PBCs offer the advantage of the fast Fourier transform (FFT) to compute spatial gradients, and therefore, we use the pseudospectral method \citep{hussaini1987spectral} to solve the governing equation in \eqref{eq:final_eqn}. Spatial gradients in \eqref{eq:final_eqn} are computed using FFT and the solution is marched forward in time using the Runge--Kutta (RK) explicit time integration with a fixed time step $\delta t$, resulting in the following discretized equations: 
\begin{equation*}
\bm \phi_\alpha^{n+1}=\bm \phi_\alpha^n+\frac{1}{6m}\left(\bm k_{\alpha 1}+2\bm k_{\alpha 2}+2 \bm k_{\alpha 3}+\bm k_{\alpha 4}\right), \text{ where}
\end{equation*}
\begin{equation*}
\bm k_{\alpha 1}=\delta t \bm f_{\alpha}|_{\bm \phi=\bm \phi^n}, \quad
\bm k_{\alpha 2}= \delta t \bm f_{\alpha}|_{\bm \phi=\bm \phi^n+\frac{\bm k_{\alpha 1}}{2}}, \quad
\bm k_{\alpha 3}= \delta t \bm f_{\alpha}|_{\bm \phi=\bm \phi^n +\frac{\bm k_{\alpha 2}}{2}}, \quad
\bm k_{\alpha 4}= \delta t \bm f_{\alpha}|_{\bm \phi=\bm \phi^n+\bm k_{\alpha 3}}.
\end{equation*}
Here, $\phi_\alpha^n:=\phi_\alpha(\cdot, t_n)$, $\bm f_\alpha$ represents the right-hand-side of \eqref{eq:final_eqn}, and $\bm f_{\alpha}|_{\bm \phi=\bm \phi^n}$ denotes the evaluation of $\bm f_\alpha$ using $\bm \phi_\alpha^n$. The spatial derivatives in $\bm f_\alpha$ are computed using FFT. The simulation domain was discretized using a $128\times128$ grid, and $\delta t=\SI{0.1}{\sec}$ was the time step size. The spatial discretization was chosen such that the width of the interface dislocations is reasonably resolved, and the temporal discretization is fixed to ensure the numerical scheme remains stable. 

All simulations are run long enough to ensure the elastic and the vdW energies converge. Since $\bm \phi_\alpha^0 \equiv \bm 0$, the elastic energy at $t=0$ is zero and the vdW energy is the only contributor to the total energy. As the simulation progresses, elastic energy increases and the vdW energy decreases, such that the total energy monotonically decreases. 

\subsection{Comparison with atomistics}
 \begin{figure}[H]
    \centering
    \subfloat[Displacement magnitude - GFK model]
    {
        \includegraphics[width=0.48\textwidth]{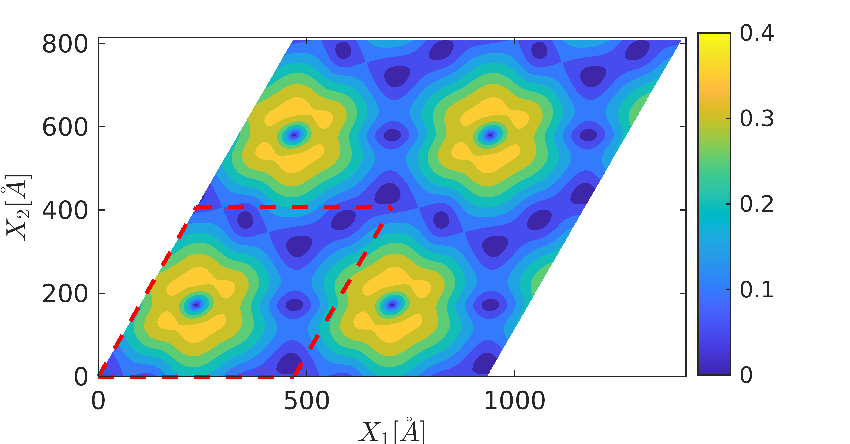}
        \label{fig:AB_cont_disp}
    }
    \subfloat[Displacement magnitude - atomistics]
    {
        \includegraphics[width=0.48\textwidth]{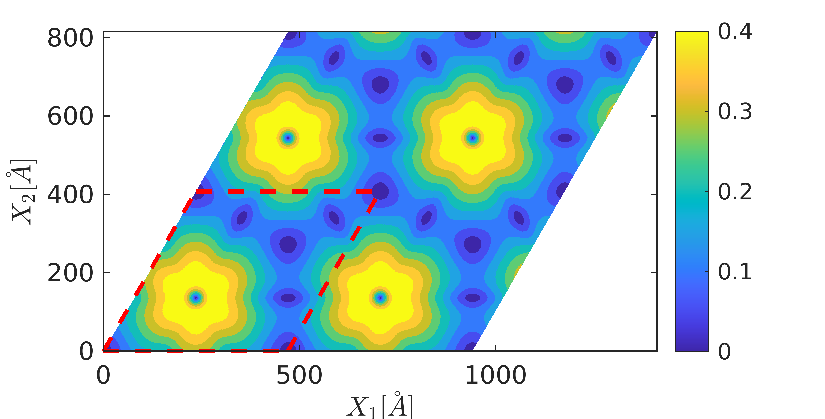}
        \label{fig:AB_lammps_disp}
    }
    \\
    \subfloat[Energy density - GFK model]
    {
        \includegraphics[width=0.48\textwidth]{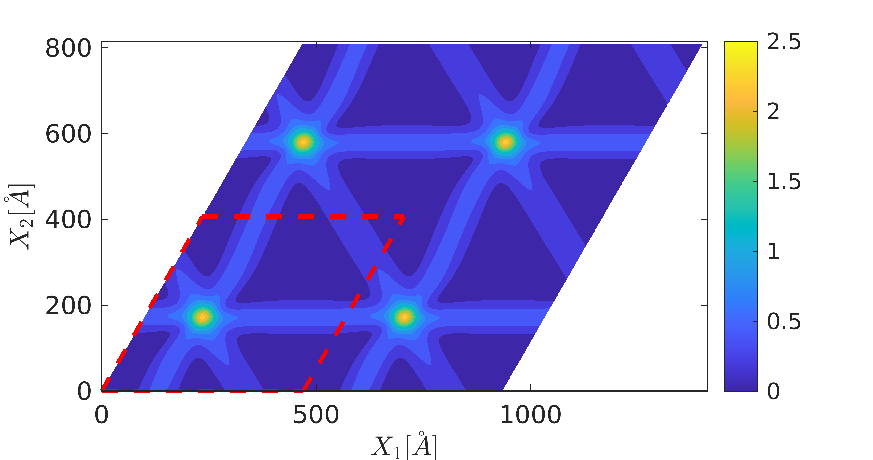}
        \label{fig:AB_cont_en}
    }
    \subfloat[Energy density - atomistics]
    {
        \includegraphics[width=0.48\textwidth]{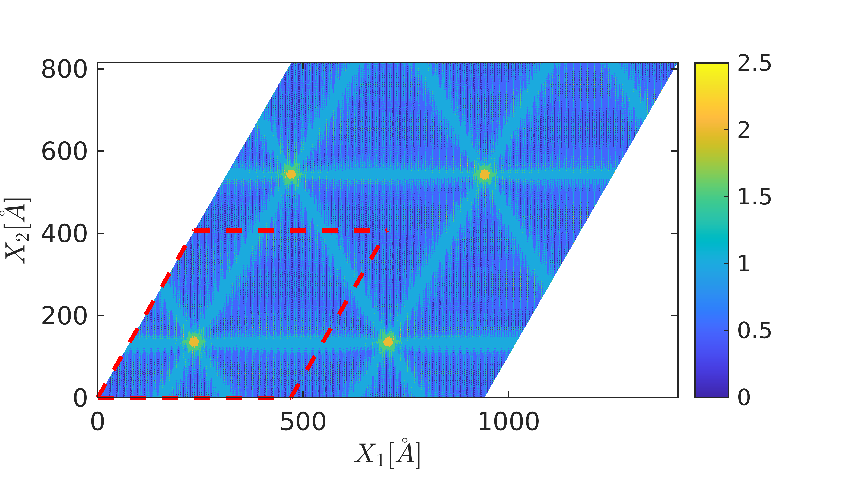}
        \label{fig:AB_lammps_en}
    }
    \caption{A comparison of structural relaxation predicted by continuum and atomistics simulations of a $\SI{0.2992634}{\degree}$ small-twist BG. The units of energy density and displacement are $\si{\meV \per \angstrom\squared}$ and $\si{\angstrom}$, respectively. The area enclosed by the red dashed line is the simulation domain.}  
    \label{fig:AB_twist_results}
\end{figure}
We will now present continuum simulations and compare them to atomistic simulations of heterodeformed BGs, discussed in \sref{sec:atomistics}. The first column of \fref{fig:AB_twist_results} shows plots of the displacement magnitude and the total energy density in a $0.2992634^\circ$ small-twist BG. They compare well with the corresponding plots from the atomistic simulation, shown in the second column. The area enclosed by the red dashed line is the simulation domain. The plots are presented on an extended domain to highlight the triangular network of dislocations.

\begin{figure}[H]
    \centering
    \subfloat[energy density in $\si{\meV \per \angstrom\squared}$]
    {
        \includegraphics[height=0.25\textwidth]{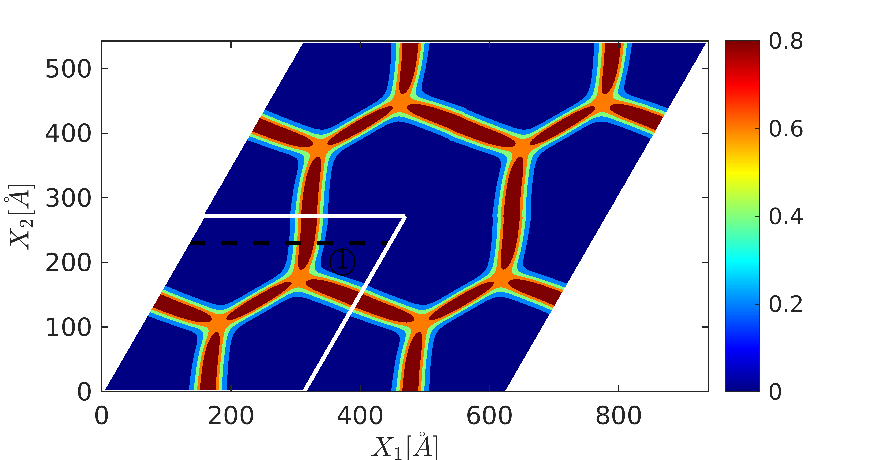}
        \label{fig:local_tw_disp_cont_old}
    }
    \subfloat[energy density in $\si{\meV \per \angstrom\squared}$]
    {
        \includegraphics[height=0.25\textwidth]{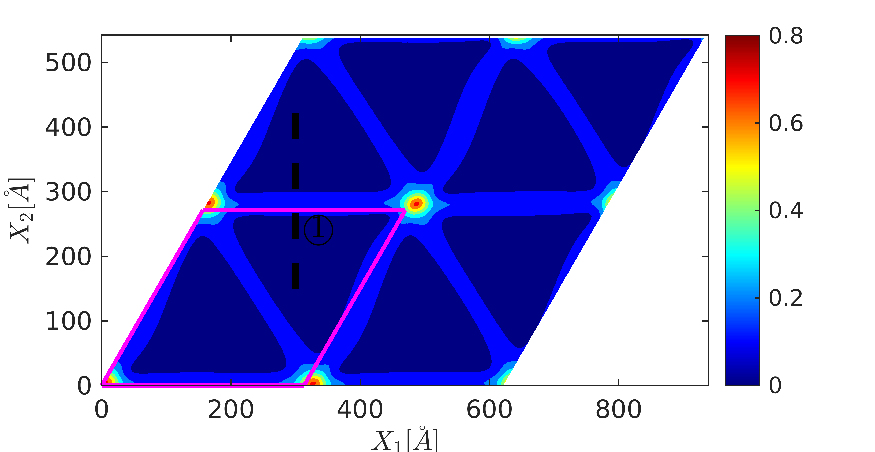}
        \label{fig:local_tw_disp_cont}
    }\\
    \subfloat[displacement $\si{\angstrom}$]
    {
        \includegraphics[height=0.25\textwidth]{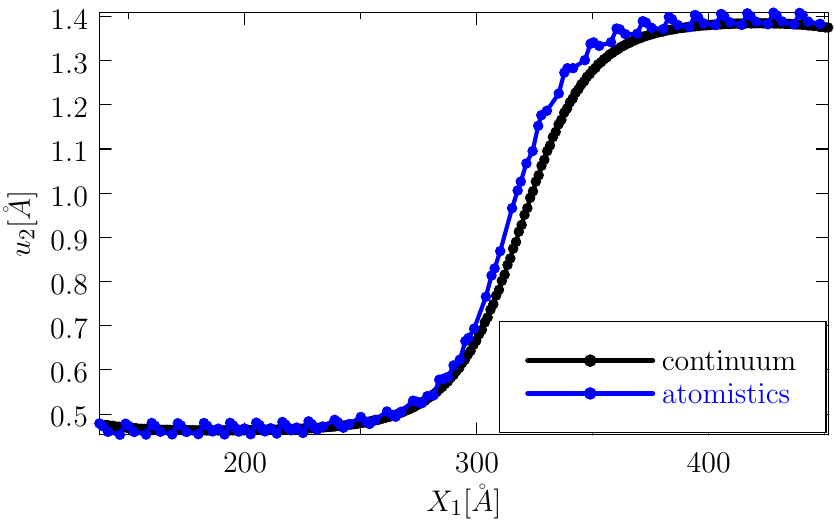}
        \label{fig:cont_atom_comp_burgers_old}
    }
    \subfloat[displacement in $\si{\angstrom}$ ]
    {
        \includegraphics[height=0.25\textwidth]{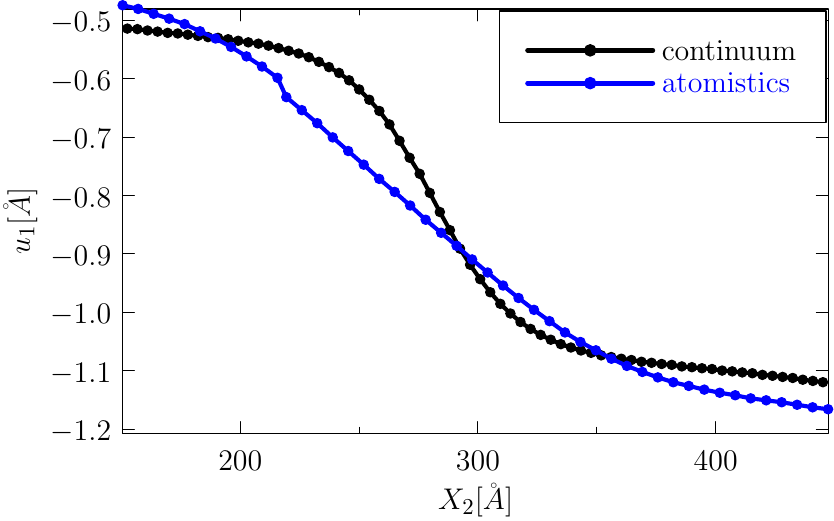}
        \label{fig:cont_atom_comp_burgers}
    }
    \caption{Structural relaxation in a $\SI{21.956865}{\angstrom}$ large-twist BG using the KC-1 and KC-2 parametrizations. The energy density plots in \psubref{fig:local_tw_disp_cont_old} and \psubref{fig:local_tw_disp_cont} highlight the  honeycomb and the triangular network of interface dislocations, corresponding to KC-1 and KC-2 models, respectively.  \psubref{fig:cont_atom_comp_burgers_old} and \psubref{fig:cont_atom_comp_burgers} compare the displacement across the dashed lines predicted by the atomistic and the FK models.}
    \label{fig:numeric_results}
    \end{figure}
Fig. \ref{fig:numeric_results} shows continuum simulation results of a $\SI{21.956865}{\degree}$ large-twist BG using vdW energy densities corresponding to the KC-1 (first column) and the KC-2 models (second column). The color density plots of the total energy density in \frefs{fig:local_tw_disp_cont_old}{fig:local_tw_disp_cont} highlight the honeycomb and triangular dislocation networks and match well with those from atomistic simulations (see \fref{fig:local_twist}). As expected, the energy density in the domain interiors is zero as they correspond to the low-energy $AB$ and $\Sigma 7$ stackings. \frefs{fig:cont_atom_comp_burgers_old}{fig:cont_atom_comp_burgers} compare the displacement jumps in the atomistic and continuum simulations.\footnote{Recall from footnote~\ref{fn:bg}, that the calculation of the Burgers vectors from the diffused displacement jump of atomistic simulations is approximate. The Burgers vector can be computed exactly from the simulations of the GFK model by considering a Burgers circuit formed by directed lines $\bm l_{\rm t}=\bm l$ and $\bm l_{\rm b}=-\bm l$ in the deformed configurations of the top and bottom lattices, respectively. The directed line $\bm l$ crosses a dislocation line and connects centers of two adjacent $\Sigma 7$/$\Sigma 1$ stackings and back. By construction, the Burgers circuit is arbitrarily thin, and its normal lies in the interface. Under this setting, the Burgers vector can be measured as
\begin{equation*}
    \int_{\bm l} (\bm H_{\rm t}^{-1} -\bm H_{\rm b}^{-1})\, d\bm l.
\end{equation*}
We have confirmed using the above equation that all simulations presented in this section recover the bicrystallography-predicted Burgers vector. 
} The displacement plots show that the core widths agree well for the KC-1 model, while the atomistic core width is relatively more diffused for the KC-2 model.

\begin{figure}[H]
    \centering
    \subfloat[displacement magnitude - atomistics]
    {
        \includegraphics[width=0.35\textwidth]{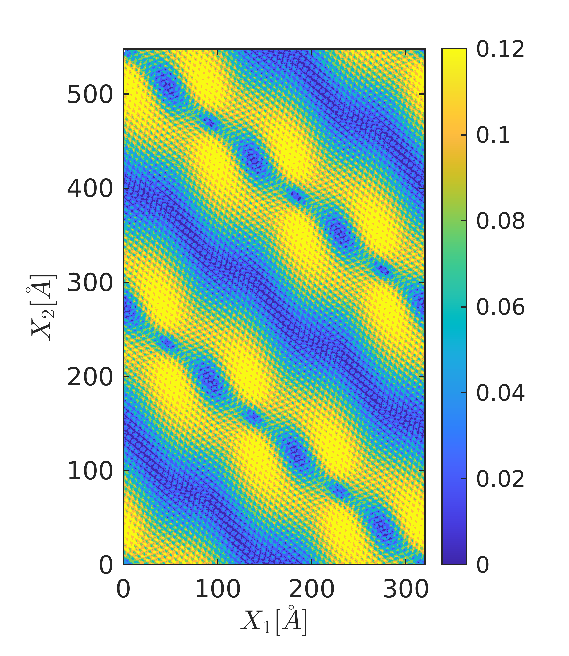}
        \label{fig:stretch_local_lammps_disp}
    }
    \subfloat[displacement magnitude - FK model]
    {
        \includegraphics[width=0.35\textwidth]{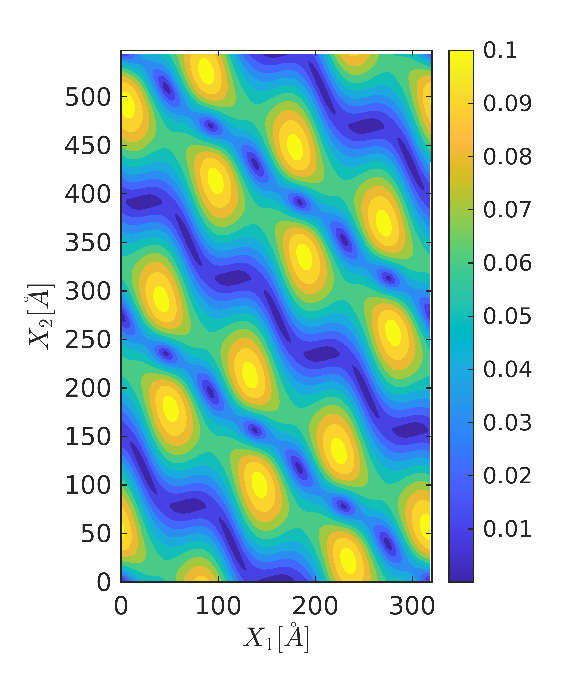}
        \label{fig:stretch_local_cont_disp}
    }
    \subfloat[energy density - FK model]
    {
        \includegraphics[width=0.37\textwidth]{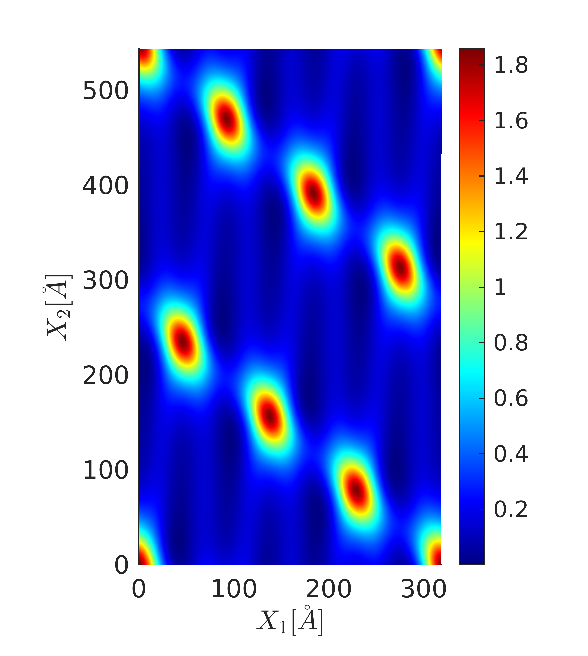}
        \label{fig:stretch_local_cont_en}
    }
    \caption{Simulation results of a heterostrained $21.786789^\circ$ large-twist BG.}
    \label{fig:local_stretch_results}
\end{figure}
\fref{fig:local_stretch_results} shows continuum simulations results of a heterostrained $21.786789^\circ$ ($\Sigma 7$) twisted BG using the vdW energy density of the KC-2 model. The displacement magnitude plotted in \fref{fig:stretch_local_cont_disp} compares well with the plot of atomistic displacement magnitude in \fref{fig:stretch_local_lammps_disp}. We note that the maximum displacement in the continuum simulation is smaller compared to that in the atomistic simulation. The dislocation network is not as conspicuous in the energy density plot in \fref{fig:stretch_local_cont_en} as it was in the earlier simulations. We attribute this feature to the diffused dislocation network noted in \fref{fig:cont_atom_comp_burgers}. We note that magnitude of the energy density shown in \fref{fig:stretch_local_cont_en} does not match with that in \frefs{fig:local_stretch_en_den_lammps_new}{fig:local_stretch_en_den_lammps_magnify} due to the atomic energy density variation of the $\Sigma 7$ configuration, as noted in \sref{sec:atomistics_large}.

\section{Summary and conclusions}
\label{sec:conclusion}
Tuning quantum mechanical properties with atomic-scale precision is at the core of scientific efforts geared towards ushering in the second quantum revolution \citep{dowling2003quantum}. New vdW materials and heterostructures are one of the key types of the novel materials that are being explored in this regard, and provide tremendous opportunities for the field of straintronics. The design and development of vdW heterostructures with tailored properties hinge on the ability to efficiently parse heterostrains and predict the properties of the resulting moir\'e superlattices. Within this context, this paper focuses on predicting atomic reconstruction in vdW heterostructures efficiently by developing a generalized Frenkel--Kontorova (GFK) model. 

Motivated by dislocations-mediated structural relaxation in a twisted BG, the development of the model was spurred by the following questions --- a) under what heterodeformations does a vdW heterostructure undergo structural relaxation?; b) is the relaxation dislocations-mediated?; and c)  how are dislocations defined in heterostructures? In our study, large twist BG serves as a surrogate for heterostructures. Noting the cusp-like local minima at an angle of $21.786789^\circ$ in the plot of interface energy versus the twist angle of a BG is a signature of defect nucleation, we hypothesized that a heterostrained $21.786789^\circ$ large-twist BG will undergo dislocations-mediated atomic reconstruction. Using atomistic simulations of $21.786789^\circ$ large-twist BG subjected to small heterotwists and heterostrains, we confirmed our hypothesis and probed the above questions. The following key observations were made in our atomistic simulations:
\begin{enumerate}
    \item Structural relaxation occurs via strain localization along a network of lines, which suggests it is dislocations-mediated. More interestingly, unlike the small twist case, the measured displacement jump/Burgers vectors were smaller than the smallest lattice vector of graphene. 
    \item Similar to the small-twist case, structural relaxation is characterized by regions of low-energy stacking interspersed by line defects. The defect-free $21.786789^\circ$ stacking is the analog of the low energy AB-stacking observed in a small-twist BG.
\end{enumerate}

To reveal the crystallographic origins of the observed dislocations, we explored the definition of an interface dislocation. Using SNF bicrystallography, which employs the Smith normal form for integer matrices, we showed that a heterointerface is translationally invariant with respect to translations in the DSCL. In other words, the GSFE of the defect-free $21.786789^\circ$ twisted BG is periodic with respect to the DSCL, which implies the Burgers vector of interface dislocations belongs to the DSCL. The GSFE from atomistics, and its periodicity inferred from SNF bicrystallography, are used to construct the interfacial energy of the GFK model. 

Inspired by crystal plasticity models, the GFK model includes three configurations --- reference, natural, and deformed. The defect-free $21.786789^\circ$ BG serves as the stress-free natural configuration, and the reference configuration is the natural configuration subjected to a uniform heterodeformation. By prescribing the constitutive law (interfacial energy and elastic energy) with respect to the natural configuration, the GFK model is rendered frame-invariant. The GFK model was used to simulate various heterodeformed BGs, and it was validated by comparing its predictions to those from atomistics. 

We conclude by emphasizing the immense potential of the GFK model to probe the enormous heterostructure-heterostrain space for correlated electron physics. Although the model is classical and focuses on structural prediction, it can serve as --- a) a workhorse for predicting structural relaxation in inhomogeneously strained heterointerfaces and b) provide a predictor for structural relaxation under uniform deformation, which can be further corrected using machine learning-based first-principles calculations \citep{pathrudkar2023electronic}. 

\section{Acknowledgements}
NCA and TA would like to acknowledge support from the National Science Foundation Grant NSF-MOMS-2239734 with S. Qidwai as the program manager. ASB and CW would like to acknowledge support through grant DE-SC0023432 funded by the U.S. Department of Energy, Office of Science. ASB and CW also acknowledge computational resource support from UCLA's Institute for Digital Research and Education (IDRE), and the National Energy Research Scientific Computing Center (NERSC awards BES-ERCAP0025205 and BES-ERCAP0025168), a DOE Office of Science User Facility supported by the Office of Science of the U.S. Department of Energy under Contract No. DE-AC02-05CH11231.

\section*{CRediT author statement}
\textbf{Md Tusher Ahmed:} Software, Validation, Formal analysis, Investigation, Data Curation, Writing-Original Draft, Visualization,
\textbf{Chenhaoyue Wang:} Software, Investigation.
\textbf{Amartya S. Banerjee:} Software, Investigation, Resources, Writing-Review \& Editing, Supervision, Funding acquisition
\textbf{Nikhil Chandra Admal:} Conceptualization, Methodology, Software, Formal analysis, Investigation, Resources, Writing-Review \& Editing, Supervision, Project Management, Funding acquisition.  

\appendix
\section{Coincidence relation between two lattices}
\label{sec:append_Coinciden_rel}
In this section, we will develop an algorithm to enumerate heterodeformations that result in moir\'e superlattices. Let $\mathcal A$ and $\mathcal B$ denote two 2D lattices with structure matrices $\bm{A}$ and $\bm{B}$, respectively. It is easy to see that the lattices coincide on a moir\'e superlattice if and only if the transition matrix,  $\bm T:=\bm A^{-1}\bm B$, is rational. However, this condition is invariably not satisfied and a heterodeformation is required to form a moir\'e supercell. Therefore, we are interested in all distortions $\bm F$ of lattice $\mathcal A$,  such that the deformed lattice shares a moir\'e supercell with $\mathcal B$. In other words, we would like to compute all $\bm F$ such that the transition matrix  
\begin{equation}
    \bm T = \bm B^{-1} \bm F \bm A \text{ is rational.}
    \label{eqn:T}
\end{equation}
Moreover, since large elastic strain are energetically unfavorable, we are only interested in heterodeformations that lead to small stretches. Heterotwists are included in this search as they cost no elastic energy.

A general solution to \eqref{eqn:T} is given by the following theorem (see \citet{ADMAL2022}):
\begin{theorem}
\label{thm:F}
Let $\mathcal A$ and $\mathcal B$ be two 2D lattices forming a heterostructure, and $\bm F$ an in-plane linear transformation. Then, $\mathcal A \cap \bm F \mathcal B$ is a 2D moir\'e superlattice if and only if there exist lattice vectors $\bm q_1$ and $\bm r_1$ in $\mathcal B$, and $\bm q_2$ and $\bm r_2$ in $\mathcal A$ such that 
\begin{equation}
    \bm q_2 = \alpha \bm F \bm q_1, \quad
    \bm r_2 = \beta \bm F \bm r_1
    \label{eqn:q2_q1}
\end{equation}
for some rationals $\alpha$ and $\beta$. Moreover, $\bm F$ is given by
\begin{equation}
    \bm F = \frac{1}{\alpha} \bm q_2 \otimes \bcal q_1 + \frac{1}{\beta} \bm r_2 \otimes \bcal r_1,
    \label{eqn:F}
\end{equation}
where $\bcal q_1$, $\bcal r_1 \in \mathcal B^*$ (dual/reciproal lattice of $\mathcal B$) such that $\bcal q_1 \cdot \bm q_1 = \bcal r_1 \cdot \bm r_1 = 1$.
\end{theorem}
An algorithm to construct $\bm F$ given in \eqref{eqn:F} can be found in \citet{ADMAL2022}, and it is implemented in \texttt{oILAB}.

\section{Smith normal form bicrystallography}
\label{sec:snf_details}
In this section, we introduce SNF bicrystallography, a framework to analyze the crystallography of heterostructures. In this paper, SNF bicrystallography is used to calculate the DSCL of a heterostructure, and plays a central role in our definition of interface dislocation, introduced in \sref{sec:snf}. We refer the reader to \citet{ADMAL2022} for a more detailed presentation.

Let $\mathcal A$ and $\mathcal B$ denote 2D lattices that form a moir\'e superlattice. From \ref{sec:append_Coinciden_rel}, we know that the transition matrix $\bm{T}=\bm{A^{-1}}\bm{B}$ is rational. Therefore, $\bm{T}$ can be expressed as 
\begin{equation}
    \bm{T}=\frac{\bm{P}}{\mu},
    \label{eqn:P}
\end{equation}
where $\mu$ is an integer, and $\bm P$ is an integer matrix such that $\mu$ and the entries of $\bm P$ are co-prime. Using the Smith normal form for integer matrices, $\bm P$ can be multiplicatively decomposed as 
\begin{equation}
    \bm P=\bm U\bm\Delta\bm V^{-1}, 
    \label{eqn:snf}
\end{equation}
where $\bm U$ and $\bm V$ are unimodular matrices, and $\bm \Delta=\textrm{diag}(\delta_1,\delta_2)$ is a diagonal matrix and $\delta_1=\gcd(\bm P)$. Substituting \eqref{eqn:snf} into \eqref{eqn:P} and rearranging, we have 

\begin{equation}
 \mu \Bp = \Ap \bm \Delta, \text{ where}
 \label{SmithT}
 \end{equation}
\begin{equation}
    \Ap=\bm A\bm U, \text{ and } \Bp=\bm B\bm V. 
    \label{eqn:ApBp}
 \end{equation}
Since $\bm U$ and $\bm V$ are unimodular, the matrices  $\Ap$ and $\Bp$ qualify as new structure matrices of lattices $\mathcal{A}$ and $\mathcal{B}$, respectively. In other words, the columns $\{\ap_i\}$ and $\{\bp_i\}$ of $\Ap$ and $\Bp$ are the new bases of the two lattices. Since $\bm \Delta$ is a diagonal matrix,  Eq.~\eqref{SmithT} reads as,
\begin{align}
    \mu \bp_i= \delta_{i}\ap_i && \text{(no summation over  i).}
    \label{SmithTcol}
\end{align}
\eqref{SmithTcol} implies the new bases are parallel and coincide on the CSL $\mathcal C$ with basis vectors 
\begin{align}
    \cp_i=\frac{\mu}{\gcd(\mu,\delta_{i})}\bp_i=\frac{\delta_{i}}{\gcd(\mu,\delta_{i})}\ap_i\, .
    \label{eqn:cp}
\end{align}
The basis vectors $\{\cp_1,\cp_2\}$ of $\mathcal C$ can be collected in a structure matrix as
\begin{align}
    \Cp = \Bp \bm N =\Ap \bm M\, . 
    \label{CSLstructMatrix}
\end{align}
where 
\begin{align}
\bm M =\text{diag}\left(\frac{\delta_{i}}{\gcd(\mu,\delta_{i})}\right) &&\text{and}&&
\bm N =\text{diag}\left(\frac{\mu}{\gcd(\mu,\delta_{i})}\right)\, ,
\label{MNmatrices}
\end{align}
are auxiliary diagonal matrices satisfying the relation $\mu \bm M =\bm \Delta \bm N$.
            
The DSCL, denoted as $\mathcal D$, is the smallest lattice that contains lattices $\mathcal A$ and $\mathcal B$. The basis vectors $\{\dpa_1,\dpa_2\}$ of $\mathcal D$ are given by
\begin{align}
\bm \dpa_i=\frac{\gcd(\mu,\delta_{i})}{\delta_{i}}\bp_i=\frac{\gcd(\mu,\delta_{i})}{\mu}\ap_i\, .
\label{eqn:dp}
\end{align}
\eqref{MNmatrices} and \eqref{eqn:dp} imply the structure matrix of $\mathcal D$ satisfies 
\begin{align}
 \Dp = \Bp \bm M^{-1}=\Ap \bm N^{-1} \, .
\label{DSCLstructMatrix}
\end{align}
\eqref{CSLstructMatrix} and \eqref{DSCLstructMatrix} imply the ratios of areas of primitive unit cells 
\begin{equation}
    \Sigma_{\mathcal A} := \frac{\det \Cp}{\det \Ap} = \det \bm M,\quad
    \Sigma_{\mathcal B} := \frac{\det \Cp}{\det \Bp} = \det \bm N
\end{equation}
are integers, and $\det(\bm C) \, \det(\bm D) =  \det(\bm A) \, \det(\bm B)$.

The parallel bases for $\mathcal A$, $\mathcal B$, $\mathcal C$, and $\mathcal D$ highlight an interesting analogy with the notions of least common multiple (lcm) and greatest common divisor (gcd) of integers. The CSL and the DSCL may be interpreted as the lcm and the gcd of lattices $\mathcal A$ and $\mathcal B$, respectively.

We will now demonstrate the application of SNF bicrystallography to the $21.786789^\circ$ twisted BG. Let $\mathcal A$ represent the hexagonal lattice of graphene with structure matrix
\begin{equation*}
    \bm A = \frac{a}{2} \begin{bmatrix}
    0 &  -\sqrt{3}\\
    2 & -1
    \end{bmatrix}, 
\end{equation*}
where the lattice constant of graphene is assumed to be $a=2.46 \AA$.
Let $\mathcal B$ represent another hexagonal lattice twisted anti-clockwise relative to $\mathcal A$, i.e. 
\begin{align*}
    \bm B = \bm R_{\theta} \bm A\, .
\end{align*}
where $\theta = 21.786789^\circ$ guarantees a coincidence between the lattices with a rational transition matrix 
\begin{equation}
    \bm T = \bm A^{-1} \bm B = \frac{1}{7} \begin{bmatrix}
      5 & -8 \\
     8 & -3 
    \end{bmatrix},
    \label{eqn:square}
\end{equation}
The SNF of the integer matrix in \eqref{eqn:square} yields 
\begin{equation*}
    \bm\Delta =\text{diag}(1,49), \quad
\bm U = \begin{bmatrix}
 19 & 1 \\
  1 & 0 
\end{bmatrix},\quad
\bm V = \begin{bmatrix}
 -1 & -3 \\
 -3 & -8 
\end{bmatrix}, \quad
\bm M =\text{diag}(1,7), \quad 
\bm N=\text{diag}(7,1).
\end{equation*}
The basis vectors of the CSL and the DSCL can be obtained using \eqref{CSLstructMatrix} and \eqref{DSCLstructMatrix}. The basis vectors of $\mathcal C$ are used to define the periodic box of our atomistic and continuum simulations, while those of $\mathcal D$ are used to identify the Burgers vectors of interface dislocations. 
While the parallel bases are valuable in proving statements such as \thref{thm:csl_shift}, the corresponding structure matrices are typically ill-conditioned. Therefore, we resort to lattice reduction algorithms to obtain reduced bases from the parallel bases.

\section{Computational details of density functional theory calculations of the GSFE of $21.786789^\circ$ twisted bilayer graphene}
\label{sec:dft}
First-principles calculations using Density Functional Theory (DFT) were performed to calculate the GSFE of $21.786789^\circ$ twisted bilayer graphene (TBG), via the Quantum Espresso package \citep{giannozziQUANTUMESPRESSOModular2009}. Projector augmented wave (PAW) type pseudopotentials \citep{blochlProjectorAugmentedWave2002}, along with the generalized gradient approximation (Perdew-Burke-Ernzerhof functional \citep{perdewPerdewBurkeErnzerhof1998}) and Grimme’s density functional dispersion correction (DFT-D2) \citep{grimmeEffectDampingFunction2011} were employed to calculate the GSFE landscape of BG systems.  These choices allowed us to balance computational accuracy and efficiency, and are also consistent with earlier literature \citep{zhouVanWaalsBilayer2015}. The interlayer spacing was set to $2.46\si{\angstrom}$, inducing $26\%$ out-of-plane compression in the BG system.  In-plane periodic boundary conditions, and out-of-plane isolated system conditions were adopted to simulate the two-dimensional nature of the system. The values of the planewave cutoff energy ($E_{\rm cut}=70$ Rydbergs) and k-point mesh parameters were optimized to ensure the total energy of systems converged to $0.001\;\si{\eV}$ in all calculations. The GSFE calculation was conducted by taking displacement on a $15\times 15 \times 1$ mesh of two lattice vectors, $\bm b_1 = -0.929792342 \, \bm e_1$, and $\bm b_2=-0.464896163 \,  \bm e_1 +0.805223601\, \bm e_2$.

\section{Comparison with the small-twist BG model of \citet{Koshino_2017}}
In this section, we will specialize the GFK model to small-twist homostructures by invoking isotropic linear elasticity to recover the model of \citet{Koshino_2017}, which was developed to model atomic reconstruction in small-twist BG.

The Frenkel--Kontorova model of \citet{Koshino_2017} for small-twist BG was formulated in terms of a single unknown field $\um$, which denotes the difference in the displacement fields of the top and bottom lattices. We will now show that under small deformations, the \citet{Koshino_2017} model can be recovered from the GFK model of \sref{sec:continuum}.

Introducing the displacement variables $\bm u_\alpha:= \bm \phi_\alpha(\bm X,t)-\bm X$, and $\bm u_{\pm}=\ut \pm \ub$, we first express $\mathcal E$ as a functional of $\up$ and $\um$ under the assumption of small deformation. Notice that we are using the variable $\bm X$, as opposed to $\bm X_\alpha$, since the two reference configurations coincide under PBCs. Beginning with $\mathcal E_{\rm el}$, we invoke linear elasticity by writing the elastic energy density as
\begin{equation}
    e_{\rm el}(\bm \epsilon_\alpha) = \frac{1}{2} \mathbb C \bm\epsilon_\alpha \cdot \bm \epsilon_\alpha= \lambda (\mathrm{tr}\, \bm \epsilon_\alpha)^2 + 2\mu \bm \epsilon_\alpha \cdot \bm \epsilon_\alpha,
\end{equation}
where the Lagrangian strain in \eqref{eqn:elasticEnergy} has been replaced by the infinitesimal strain $\bm \epsilon_\alpha=(\nabla \bm u_\alpha + \nabla \bm u_\alpha^{\rm T})/2$, and $\mathbb C$ is the fourth-order isotropic elasticity tensor with lam\'e constants $\lambda$ and $\mu$.
Under this setting, $\mathcal E_{\rm el}$ can be recast as a functional of $\up$ and $\um$ as follows:
\begin{equation}
    \mathcal E_{\rm el}[\up,\um]=\frac{1}{4}  \int_{\Omega_{\rm ref}} (\mathbb C \bm \epsilon_+ \cdot \bm \epsilon_+ + \mathbb C \bm \epsilon_- \cdot \bm \epsilon_-) \, d\bm X.
\end{equation}
where $\epsilon_{\pm}:= (\nabla \bm u_{\pm} + \nabla \bm u_{\pm}^{\rm T})/2$. Next, assuming a) $\det \bm K_\alpha \approx \det \bm F_\alpha \approx 1$ and b) $\bm H_\alpha \approx \bm I$, the vdW energy can be expressed as a functional exclusively of $\um$:
\begin{equation}
    \mathcal E_{\rm vdW}[\um] 
    = \int_{\Omega^{\rm ref}} 
    e_{\rm vdW}(\bm r(\bm X)) \, d\bm X,
    \label{eqn:vdWEnergy_Koshino}
\end{equation}
where $\bm r(\bm X)$, given by \eqref{eqn:r}, is written as $\bm r(\bm X)= (\Kt-\Kb)\bm X - \um(\bm X)$ since $\phit-\phib=\ut-\ub=\um$. 

Since $\mathcal E_{\rm vdW}$ in \eqref{eqn:vdWEnergy_Koshino} is independent of $\up$, it is easy to see that minimizing the total energy functional $\mathcal E[\up,\um]$ results in $\up\equiv 0$, which allowed \citet{Koshino_2017} to cast their total energy functional in the single variable $\um$:
\begin{equation}
    \mathcal E[\um]= \int_{\Omega_{\rm ref}}
    \left (
        \frac{1}{4}\mathbb C \bm \epsilon_- \cdot \bm \epsilon_- +
        e_{\rm vdW}(\bm r)
    \right )\, d\bm X.
\end{equation}

\bibliography{apssamp}

\providecommand{\noopsort}[1]{}\providecommand{\singleletter}[1]{#1}%
\begin{thebibliography}{64}
\providecommand{\natexlab}[1]{#1}
\providecommand{\url}[1]{\texttt{#1}}
\expandafter\ifx\csname urlstyle\endcsname\relax
  \providecommand{\doi}[1]{doi: #1}\else
  \providecommand{\doi}{doi: \begingroup \urlstyle{rm}\Url}\fi

\bibitem[Admal et~al.(2018)Admal, Po, and Marian]{ADMAL2018}
Nikhil~Chandra Admal, Giacomo Po, and Jaime Marian.
\newblock A unified framework for polycrystal plasticity with grain boundary
  evolution.
\newblock \emph{International Journal of Plasticity}, 106:\penalty0 1--30,
  2018.
\newblock ISSN 0749-6419.
\newblock \doi{https://doi.org/10.1016/j.ijplas.2018.01.014}.
\newblock URL
  \url{https://www.sciencedirect.com/science/article/pii/S0749641917305557}.

\bibitem[Admal et~al.(2022)Admal, Ahmed, Martinez, and Po]{ADMAL2022}
Nikhil~Chandra Admal, Tusher Ahmed, Enrique Martinez, and Giacomo Po.
\newblock Interface dislocations and grain boundary disconnections using smith
  normal bicrystallography.
\newblock \emph{Acta Materialia}, 240:\penalty0 118340, 2022.
\newblock ISSN 1359-6454.
\newblock \doi{https://doi.org/10.1016/j.actamat.2022.118340}.
\newblock URL
  \url{https://www.sciencedirect.com/science/article/pii/S1359645422007194}.

\bibitem[Alden et~al.(2013)Alden, Tsen, Huang, Hovden, Brown, Park, Muller, and
  McEuen]{alden2013strain}
Jonathan~S Alden, Adam~W Tsen, Pinshane~Y Huang, Robert Hovden, Lola Brown,
  Jiwoong Park, David~A Muller, and Paul~L McEuen.
\newblock Strain solitons and topological defects in bilayer graphene.
\newblock \emph{Proceedings of the National Academy of Sciences}, 110\penalty0
  (28):\penalty0 11256--11260, 2013.

\bibitem[Annevelink et~al.(2020)Annevelink, Johnson, and
  Ertekin]{Annevelink_2020}
Emil Annevelink, Harley~T. Johnson, and Elif Ertekin.
\newblock Topologically derived dislocation theory for twist and stretch
  moir\'e superlattices in bilayer graphene.
\newblock \emph{Phys. Rev. B}, 102:\penalty0 184107, Nov 2020.
\newblock \doi{10.1103/PhysRevB.102.184107}.
\newblock URL \url{https://link.aps.org/doi/10.1103/PhysRevB.102.184107}.

\bibitem[Annevelink et~al.(2021)Annevelink, Wang, Dong, Johnson, and
  Pochet]{annevelink2021moire}
Emil Annevelink, Zhu-Jun Wang, Guocai Dong, Harley~T Johnson, and Pascal
  Pochet.
\newblock A moir{\'e} theory for probing grain boundary structure in graphene.
\newblock \emph{Acta Materialia}, 217:\penalty0 117156, 2021.

\bibitem[Balluffi et~al.(1982)Balluffi, Brokman, and King]{Balluffi_1982}
R~W Balluffi, A~Brokman, and A~H King.
\newblock Csl/dsc lattice model for general crystal boundaries and their line
  defects.
\newblock \emph{Acta Metall.; (United States)}, 30:8, 8 1982.
\newblock \doi{10.1016/0001-6160(82)90166-3}.
\newblock URL \url{https://www.osti.gov/biblio/5907589}.

\bibitem[Basov et~al.(2017)Basov, Averitt, and Hsieh]{basov2017towards}
DN~Basov, RD~Averitt, and D~Hsieh.
\newblock Towards properties on demand in quantum materials.
\newblock \emph{Nature materials}, 16\penalty0 (11):\penalty0 1077--1088, 2017.

\bibitem[Bistritzer and MacDonald(2011)]{Macdonald_2011}
Rafi Bistritzer and Allan~H. MacDonald.
\newblock Moiré bands in twisted double-layer graphene.
\newblock \emph{Proceedings of the National Academy of Sciences}, 108\penalty0
  (30):\penalty0 12233--12237, 2011.
\newblock \doi{10.1073/pnas.1108174108}.
\newblock URL \url{https://www.pnas.org/doi/abs/10.1073/pnas.1108174108}.

\bibitem[Bitzek et~al.(2006)Bitzek, Koskinen, G\"ahler, Moseler, and
  Gumbsch]{bitzek2006fire}
Erik Bitzek, Pekka Koskinen, Franz G\"ahler, Michael Moseler, and Peter
  Gumbsch.
\newblock Structural relaxation made simple.
\newblock \emph{Phys. Rev. Lett.}, 97:\penalty0 170201, Oct 2006.
\newblock \doi{10.1103/PhysRevLett.97.170201}.
\newblock URL \url{https://link.aps.org/doi/10.1103/PhysRevLett.97.170201}.

\bibitem[Bl{\"o}chl et~al.(2002)Bl{\"o}chl, F{\"o}rst, and
  Schimpl]{blochlProjectorAugmentedWave2002}
Peter~E. Bl{\"o}chl, Clemens~J. F{\"o}rst, and Johannes Schimpl.
\newblock The {{Projector Augmented Wave Method}}: Ab-initio molecular dynamics
  with full wave functions, July 2002.

\bibitem[Brenner et~al.(2002)Brenner, Shenderova, Harrison, Stuart, Ni, and
  Sinnott]{Rebo_Brenner_2002}
Donald~W Brenner, Olga~A Shenderova, Judith~A Harrison, Steven~J Stuart, Boris
  Ni, and Susan~B Sinnott.
\newblock A second-generation reactive empirical bond order (rebo) potential
  energy expression for hydrocarbons.
\newblock \emph{Journal of Physics: Condensed Matter}, 14\penalty0
  (4):\penalty0 783, jan 2002.
\newblock \doi{10.1088/0953-8984/14/4/312}.
\newblock URL \url{https://dx.doi.org/10.1088/0953-8984/14/4/312}.

\bibitem[Cao et~al.(2018)Cao, Fatemi, Fang, Watanabe, Taniguchi, Kaxiras, and
  Jarillo-Herrero]{cao_unconventional_2018}
Yuan Cao, Valla Fatemi, Shiang Fang, Kenji Watanabe, Takashi Taniguchi,
  Efthimios Kaxiras, and Pablo Jarillo-Herrero.
\newblock Unconventional superconductivity in magic-angle graphene
  superlattices.
\newblock \emph{Nature}, 556\penalty0 (7699):\penalty0 43--50, April 2018.
\newblock ISSN 1476-4687.
\newblock \doi{10.1038/nature26160}.
\newblock URL \url{https://www.nature.com/articles/nature26160}.

\bibitem[Cao et~al.(2020)Cao, Rodan-Legrain, Rubies-Bigorda, Park, Watanabe,
  Taniguchi, and Jarillo-Herrero]{cao_2020}
Yuan Cao, Daniel Rodan-Legrain, Oriol Rubies-Bigorda, Jeong~Min Park, Kenji
  Watanabe, Takashi Taniguchi, and Pablo Jarillo-Herrero.
\newblock Tunable correlated states and spin-polarized phases in twisted
  bilayer–bilayer graphene.
\newblock \emph{Nature}, 583\penalty0 (7815):\penalty0 215–220, 2020.
\newblock \doi{10.1038/s41586-020-2260-6}.

\bibitem[Cao et~al.(2021)Cao, Rodan-Legrain, Park, Yuan, Watanabe, Taniguchi,
  Fernandes, Fu, and Jarillo-Herrero]{Cao2021}
Yuan Cao, Daniel Rodan-Legrain, Jeong~Min Park, Noah F.~Q. Yuan, Kenji
  Watanabe, Takashi Taniguchi, Rafael~M. Fernandes, Liang Fu, and Pablo
  Jarillo-Herrero.
\newblock Nematicity and competing orders in superconducting magic-angle
  graphene.
\newblock \emph{Science}, 372\penalty0 (6539):\penalty0 264--271, 2021.
\newblock \doi{10.1126/science.abc2836}.
\newblock URL \url{https://www.science.org/doi/abs/10.1126/science.abc2836}.

\bibitem[Carr et~al.(2018{\natexlab{a}})Carr, Fang, Jarillo-Herrero, and
  Kaxiras]{Carr2018}
Stephen Carr, Shiang Fang, Pablo Jarillo-Herrero, and Efthimios Kaxiras.
\newblock Pressure dependence of the magic twist angle in graphene
  superlattices.
\newblock \emph{Phys. Rev. B}, 98:\penalty0 085144, Aug 2018{\natexlab{a}}.
\newblock \doi{10.1103/PhysRevB.98.085144}.
\newblock URL \url{https://link.aps.org/doi/10.1103/PhysRevB.98.085144}.

\bibitem[Carr et~al.(2018{\natexlab{b}})Carr, Massatt, Torrisi, Cazeaux,
  Luskin, and Kaxiras]{Carr_2018_continuum}
Stephen Carr, Daniel Massatt, Steven~B. Torrisi, Paul Cazeaux, Mitchell Luskin,
  and Efthimios Kaxiras.
\newblock Relaxation and domain formation in incommensurate two-dimensional
  heterostructures.
\newblock \emph{Phys. Rev. B}, 98:\penalty0 224102, Dec 2018{\natexlab{b}}.
\newblock \doi{10.1103/PhysRevB.98.224102}.
\newblock URL \url{https://link.aps.org/doi/10.1103/PhysRevB.98.224102}.

\bibitem[Cazeaux et~al.(2020)Cazeaux, Luskin, and Massatt]{cazeaux2020energy}
Paul Cazeaux, Mitchell Luskin, and Daniel Massatt.
\newblock Energy minimization of two dimensional incommensurate
  heterostructures.
\newblock \emph{Archive for Rational Mechanics and Analysis}, 235\penalty0
  (2):\penalty0 1289--1325, 2020.

\bibitem[Cazeaux et~al.(2023)Cazeaux, Clark, Engelke, Kim, and
  Luskin]{cazeaux2023relaxation}
Paul Cazeaux, Drake Clark, Rebecca Engelke, Philip Kim, and Mitchell Luskin.
\newblock Relaxation and domain wall structure of bilayer moire systems.
\newblock \emph{Journal of Elasticity}, pages 1--24, 2023.

\bibitem[Cheng et~al.(2023)Cheng, Yuan, Zhou, and Zhou]{cheng2023moire}
Xiaobian Cheng, Jie Yuan, Benhu Zhou, and Benliang Zhou.
\newblock The moir{\'e} pattern rule of the twisted bilayer graphene and its
  electronic property under a strain.
\newblock \emph{The European Physical Journal Plus}, 138\penalty0 (1):\penalty0
  1--6, 2023.

\bibitem[Chittari et~al.(2018)Chittari, Leconte, Javvaji, and
  Jung]{Chittari_2019}
Bheema~Lingam Chittari, Nicolas Leconte, Srivani Javvaji, and Jeil Jung.
\newblock Pressure induced compression of flatbands in twisted bilayer
  graphene.
\newblock \emph{Electronic Structure}, 1\penalty0 (1):\penalty0 015001, nov
  2018.
\newblock \doi{10.1088/2516-1075/aaead3}.
\newblock URL \url{https://dx.doi.org/10.1088/2516-1075/aaead3}.

\bibitem[Clayton(2010)]{clayton2010}
John~D Clayton.
\newblock \emph{Nonlinear mechanics of crystals}, volume 177.
\newblock Springer Science \& Business Media, 2010.

\bibitem[Dai et~al.(2016)Dai, Xiang, and Srolovitz]{srolovitz_2016}
Shuyang Dai, Yang Xiang, and David~J. Srolovitz.
\newblock Twisted bilayer graphene: Moiré with a twist.
\newblock \emph{Nano Letters}, 16\penalty0 (9):\penalty0 5923–5927, 2016.
\newblock \doi{10.1021/acs.nanolett.6b02870}.

\bibitem[Das et~al.(2016)Das, Bhattacharyya, Mu\~noz, and
  Singh]{Das_2016_compression}
Deya Das, Swastibrata Bhattacharyya, Enrique Mu\~noz, and Abhishek~K. Singh.
\newblock Strain-induced chiral symmetry breaking leads to large dirac cone
  splitting in graphene/graphane heterostructure.
\newblock \emph{Phys. Rev. B}, 94:\penalty0 115438, Sep 2016.
\newblock \doi{10.1103/PhysRevB.94.115438}.
\newblock URL \url{https://link.aps.org/doi/10.1103/PhysRevB.94.115438}.

\bibitem[Dowling and Milburn(2003)]{dowling2003quantum}
Jonathan~P Dowling and Gerard~J Milburn.
\newblock Quantum technology: the second quantum revolution.
\newblock \emph{Philosophical Transactions of the Royal Society of London.
  Series A: Mathematical, Physical and Engineering Sciences}, 361\penalty0
  (1809):\penalty0 1655--1674, 2003.

\bibitem[Gargiulo and Yazyev(2017)]{Gargiulo_2018}
Fernando Gargiulo and Oleg~V Yazyev.
\newblock Structural and electronic transformation in low-angle twisted bilayer
  graphene.
\newblock \emph{2D Materials}, 5\penalty0 (1):\penalty0 015019, nov 2017.
\newblock \doi{10.1088/2053-1583/aa9640}.
\newblock URL \url{https://dx.doi.org/10.1088/2053-1583/aa9640}.

\bibitem[Giannozzi et~al.(2009)Giannozzi, Baroni, Bonini, Calandra, Car,
  Cavazzoni, Ceresoli, Chiarotti, Cococcioni, Dabo, Corso, de~Gironcoli,
  Fabris, Fratesi, Gebauer, Gerstmann, Gougoussis, Kokalj, Lazzeri,
  {Martin-Samos}, Marzari, Mauri, Mazzarello, Paolini, Pasquarello, Paulatto,
  Sbraccia, Scandolo, Sclauzero, Seitsonen, Smogunov, Umari, and
  Wentzcovitch]{giannozziQUANTUMESPRESSOModular2009}
Paolo Giannozzi, Stefano Baroni, Nicola Bonini, Matteo Calandra, Roberto Car,
  Carlo Cavazzoni, Davide Ceresoli, Guido~L. Chiarotti, Matteo Cococcioni,
  Ismaila Dabo, Andrea~Dal Corso, Stefano de~Gironcoli, Stefano Fabris, Guido
  Fratesi, Ralph Gebauer, Uwe Gerstmann, Christos Gougoussis, Anton Kokalj,
  Michele Lazzeri, Layla {Martin-Samos}, Nicola Marzari, Francesco Mauri,
  Riccardo Mazzarello, Stefano Paolini, Alfredo Pasquarello, Lorenzo Paulatto,
  Carlo Sbraccia, Sandro Scandolo, Gabriele Sclauzero, Ari~P. Seitsonen,
  Alexander Smogunov, Paolo Umari, and Renata~M. Wentzcovitch.
\newblock {{QUANTUM ESPRESSO}}: A modular and open-source software project for
  quantum simulations of materials.
\newblock \emph{J. Phys.: Condens. Matter}, 21\penalty0 (39):\penalty0 395502,
  September 2009.
\newblock ISSN 0953-8984.
\newblock \doi{10.1088/0953-8984/21/39/395502}.

\bibitem[Grimme et~al.(2011)Grimme, Ehrlich, and
  Goerigk]{grimmeEffectDampingFunction2011}
Stefan Grimme, Stephan Ehrlich, and Lars Goerigk.
\newblock Effect of the damping function in dispersion corrected density
  functional theory.
\newblock \emph{Journal of Computational Chemistry}, 32\penalty0 (7):\penalty0
  1456--1465, 2011.
\newblock ISSN 1096-987X.
\newblock \doi{10.1002/jcc.21759}.

\bibitem[Grimmer et~al.(1974)Grimmer, Bollmann, and Warrington]{bollman_1974}
H~Grimmer, WT~Bollmann, and DH~Warrington.
\newblock Coincidence-site lattices and complete pattern-shift in cubic
  crystals.
\newblock \emph{Acta Crystallographica Section A: Crystal Physics, Diffraction,
  Theoretical and General Crystallography}, 30\penalty0 (2):\penalty0 197--207,
  1974.

\bibitem[Hamer et~al.(2022)Hamer, Giampietri, Kandyba, Genuzio, Menteş,
  Locatelli, Gorbachev, Barinov, and Mucha-Kruczyński]{nano2022localmoire}
Matthew~J. Hamer, Alessio Giampietri, Viktor Kandyba, Francesca Genuzio,
  Tevfik~O. Menteş, Andrea Locatelli, Roman~V. Gorbachev, Alexei Barinov, and
  Marcin Mucha-Kruczyński.
\newblock Moiré superlattice effects and band structure evolution in
  near-30-degree twisted bilayer graphene.
\newblock \emph{ACS Nano}, 16\penalty0 (2):\penalty0 1954--1962, 2022.
\newblock \doi{10.1021/acsnano.1c06439}.
\newblock URL \url{https://doi.org/10.1021/acsnano.1c06439}.
\newblock PMID: 35073479.

\bibitem[He and Admal(2021)]{Junyan_2021}
Junyan He and Nikhil~Chandra Admal.
\newblock Polycrystal plasticity with grain boundary evolution: A numerically
  efficient dislocation-based diffuse-interface model.
\newblock \emph{Modelling and Simulation in Materials Science and Engineering},
  30, 10 2021.
\newblock \doi{10.1088/1361-651X/ac2f84}.

\bibitem[Hussaini and Zang(1987)]{hussaini1987spectral}
M~Yousuff Hussaini and Thomas~A Zang.
\newblock Spectral methods in fluid dynamics.
\newblock \emph{Annual review of fluid mechanics}, 19\penalty0 (1):\penalty0
  339--367, 1987.

\bibitem[Ishikawa et~al.(2016)Ishikawa, Lugg, Inoue, Sawada, Taniguchi,
  Shibata, and Ikuhara]{ishikawa2016interfacial}
Ryo Ishikawa, Nathan~R Lugg, Kazutoshi Inoue, Hidetaka Sawada, Takashi
  Taniguchi, Naoya Shibata, and Yuichi Ikuhara.
\newblock Interfacial atomic structure of twisted few-layer graphene.
\newblock \emph{Scientific reports}, 6\penalty0 (1):\penalty0 21273, 2016.

\bibitem[Jin et~al.(2019)Jin, Regan, Yan, Iqbal Bakti~Utama, Wang, Zhao, Qin,
  Yang, Zheng, Shi, Watanabe, Taniguchi, Tongay, Zettl, and Wang]{Jin_2019}
Chenhao Jin, Emma~C. Regan, Aiming Yan, M.~Iqbal Bakti~Utama, Danqing Wang,
  Sihan Zhao, Ying Qin, Sijie Yang, Zhiren Zheng, Shenyang Shi, Kenji Watanabe,
  Takashi Taniguchi, Sefaattin Tongay, Alex Zettl, and Feng Wang.
\newblock Observation of moiré excitons in wse2/ws2 heterostructure
  superlattices.
\newblock \emph{Nature (London)}, 567\penalty0 (7746), 2 2019.

\bibitem[Joshi et~al.(2022)Joshi, He, and Admal]{Himanshu_2022}
Himanshu Joshi, Junyan He, and Nikhil~Chandra Admal.
\newblock A finite deformation theory for grain boundary plasticity based on
  geometrically necessary disconnections.
\newblock \emph{Journal of the Mechanics and Physics of Solids}, 167:\penalty0
  104949, 06 2022.
\newblock \doi{10.1016/j.jmps.2022.104949}.

\bibitem[Keimer and Moore(2017)]{keimer2017physics}
B~Keimer and JE~Moore.
\newblock The physics of quantum materials.
\newblock \emph{Nature Physics}, 13\penalty0 (11):\penalty0 1045--1055, 2017.

\bibitem[Kim et~al.(2022)Kim, Haque, Hsieh, Nahid, Zarin, Jeong, So, Park, and
  Nam]{kim2022strain}
Jin~Myung Kim, Md~Farhadul Haque, Ezekiel~Y Hsieh, Shahriar~Muhammad Nahid,
  Ishrat Zarin, Kwang-Yong Jeong, Jae-Pil So, Hong-Gyu Park, and SungWoo Nam.
\newblock Strain engineering of low-dimensional materials for emerging quantum
  phenomena and functionalities.
\newblock \emph{Advanced Materials}, page 2107362, 2022.

\bibitem[Kim et~al.(2016)Kim, Yankowitz, Fallahazad, Kang, Movva, Huang,
  Larentis, Corbet, Taniguchi, Watanabe, et~al.]{kim2016van}
Kyounghwan Kim, Matthew Yankowitz, Babak Fallahazad, Sangwoo Kang, Hema~CP
  Movva, Shengqiang Huang, Stefano Larentis, Chris~M Corbet, Takashi Taniguchi,
  Kenji Watanabe, et~al.
\newblock van der waals heterostructures with high accuracy rotational
  alignment.
\newblock \emph{Nano letters}, 16\penalty0 (3):\penalty0 1989--1995, 2016.

\bibitem[Koda et~al.(2016)Koda, Bechstedt, Marques, and
  Teles]{koda2016coincidence}
Daniel~S Koda, Friedhelm Bechstedt, Marcelo Marques, and Lara~K Teles.
\newblock Coincidence lattices of 2d crystals: Heterostructure predictions and
  applications.
\newblock \emph{The Journal of Physical Chemistry C}, 120\penalty0
  (20):\penalty0 10895--10908, 2016.

\bibitem[Kolmogorov and Crespi(2005)]{kolmogorov2005kc}
Aleksey~N. Kolmogorov and Vincent~H. Crespi.
\newblock Registry-dependent interlayer potential for graphitic systems.
\newblock \emph{Phys. Rev. B}, 71:\penalty0 235415, Jun 2005.
\newblock \doi{10.1103/PhysRevB.71.235415}.
\newblock URL \url{https://link.aps.org/doi/10.1103/PhysRevB.71.235415}.

\bibitem[Kumar et~al.(2016)Kumar, Dong, and Shenoy]{kumar2016limits}
Hemant Kumar, Liang Dong, and Vivek~B Shenoy.
\newblock Limits of coherency and strain transfer in flexible 2d van der waals
  heterostructures: formation of strain solitons and interlayer debonding.
\newblock \emph{Scientific reports}, 6\penalty0 (1):\penalty0 21516, 2016.

\bibitem[Lee et~al.(2019)Lee, Park, and Choi]{lee_park_choi_2019}
Jin-Ho Lee, Soo-jeong Park, and Jeong-Woo Choi.
\newblock Electrical property of graphene and its application to
  electrochemical biosensing.
\newblock \emph{Nanomaterials}, 9\penalty0 (2):\penalty0 297, 2019.
\newblock \doi{10.3390/nano9020297}.

\bibitem[Lopes~dos Santos et~al.(2012)Lopes~dos Santos, Peres, and
  Castro~Neto]{Lopes2012}
J.~M.~B. Lopes~dos Santos, N.~M.~R. Peres, and A.~H. Castro~Neto.
\newblock Continuum model of the twisted graphene bilayer.
\newblock \emph{Phys. Rev. B}, 86:\penalty0 155449, Oct 2012.
\newblock \doi{10.1103/PhysRevB.86.155449}.
\newblock URL \url{https://link.aps.org/doi/10.1103/PhysRevB.86.155449}.

\bibitem[Miao et~al.(2021)Miao, Liang, and Cheng]{miao2021straintronics}
Feng Miao, Shi-Jun Liang, and Bin Cheng.
\newblock Straintronics with van der waals materials.
\newblock \emph{npj Quantum Materials}, 6\penalty0 (1):\penalty0 59, 2021.

\bibitem[Morovati et~al.(2022)Morovati, Xue, Liechti, and
  Huang]{morovati2022interlayer}
Vahid Morovati, Zhiming Xue, Kenneth~M Liechti, and Rui Huang.
\newblock Interlayer coupling and strain localization in small-twist-angle
  graphene flakes.
\newblock \emph{Extreme Mechanics Letters}, 55:\penalty0 101829, 2022.

\bibitem[Nam and Koshino(2017)]{Koshino_2017}
Nguyen~NT Nam and Mikito Koshino.
\newblock Lattice relaxation and energy band modulation in twisted bilayer
  graphene.
\newblock \emph{Physical Review B}, 96\penalty0 (7):\penalty0 075311, 2017.

\bibitem[Ouyang et~al.(2018)Ouyang, Mandelli, Urbakh, and
  Hod]{ouyang2018nanoserpents}
Wengen Ouyang, Davide Mandelli, Michael Urbakh, and Oded Hod.
\newblock Nanoserpents: Graphene nanoribbon motion on two-dimensional hexagonal
  materials.
\newblock \emph{Nano letters}, 18\penalty0 (9):\penalty0 6009--6016, 2018.

\bibitem[Papageorgiou et~al.(2017)Papageorgiou, Kinloch, and
  Young]{PAPAGEORGIOU201775}
Dimitrios~G. Papageorgiou, Ian~A. Kinloch, and Robert~J. Young.
\newblock Mechanical properties of graphene and graphene-based nanocomposites.
\newblock \emph{Progress in Materials Science}, 90:\penalty0 75--127, 2017.
\newblock ISSN 0079-6425.
\newblock \doi{https://doi.org/10.1016/j.pmatsci.2017.07.004}.
\newblock URL
  \url{https://www.sciencedirect.com/science/article/pii/S0079642517300968}.

\bibitem[Pathrudkar et~al.(2023)Pathrudkar, Thiagarajan, Agarwal, Banerjee, and
  Ghosh]{pathrudkar2023electronic}
Shashank Pathrudkar, Ponkrshnan Thiagarajan, Shivang Agarwal, Amartya~S
  Banerjee, and Susanta Ghosh.
\newblock Electronic structure prediction of multi-million atom systems through
  uncertainty quantification enabled transfer learning.
\newblock \emph{arXiv preprint arXiv:2308.13096}, 2023.

\bibitem[Perdew et~al.(1998)Perdew, Burke, and
  Ernzerhof]{perdewPerdewBurkeErnzerhof1998}
J.~P. Perdew, K.~Burke, and M.~Ernzerhof.
\newblock Perdew, {{Burke}}, and {{Ernzerhof Reply}}:.
\newblock \emph{Phys. Rev. Lett.}, 80\penalty0 (4):\penalty0 891--891, January
  1998.
\newblock \doi{10.1103/PhysRevLett.80.891}.

\bibitem[Plimpton(1995)]{plimpton1995fast}
Steve Plimpton.
\newblock Fast parallel algorithms for short-range molecular dynamics.
\newblock \emph{Journal of computational physics}, 117\penalty0 (1):\penalty0
  1--19, 1995.

\bibitem[Pochet et~al.(2017)Pochet, McGuigan, Coraux, and
  Johnson]{harley_disloc}
Pascal Pochet, Brian~C. McGuigan, Johann Coraux, and Harley~T. Johnson.
\newblock Toward moiré engineering in 2d materials via dislocation theory.
\newblock \emph{Applied Materials Today}, 9:\penalty0 240--250, 2017.
\newblock ISSN 2352-9407.
\newblock \doi{https://doi.org/10.1016/j.apmt.2017.07.007}.
\newblock URL
  \url{https://www.sciencedirect.com/science/article/pii/S2352940717302305}.

\bibitem[Rakib et~al.(2022)Rakib, Pochet, Ertekin, and Johnson]{tawfiq_2022}
Tawfiqur Rakib, Pascal Pochet, Elif Ertekin, and Harley~T. Johnson.
\newblock Corrugation-driven symmetry breaking in magic-angle twisted bilayer
  graphene.
\newblock \emph{Communications Physics}, 5\penalty0 (1), 10 2022.
\newblock \doi{10.1038/s42005-022-01013-y}.

\bibitem[Regan et~al.(2020)Regan, Wang, Jin, Bakti~Utama, Gao, Zhao, Zhao,
  Zhang, Yumigeta, Blei, Carlström, Watanabe, Taniguchi, Tongay, Crommie,
  Zettl, Wang, and Wei]{Regan_2020}
Emma~C. Regan, Danqing Wang, Chenhao Jin, M.~Iqbal Bakti~Utama, Beini Gao,
  Sihan Zhao, Wenyu Zhao, Zuocheng Zhang, Kentaro Yumigeta, Mark Blei, Johan~D.
  Carlström, Kenji Watanabe, Takashi Taniguchi, Sefaattin Tongay, Michael
  Crommie, Alex Zettl, Feng Wang, and Xin Wei.
\newblock Mott and generalized wigner crystal states in wse2/ws2 moiré
  superlattices.
\newblock \emph{Nature (London)}, 579\penalty0 (7799), 3 2020.
\newblock \doi{10.1038/s41586-020-2092-4}.

\bibitem[Shimazaki et~al.(2020)Shimazaki, Schwartz, Watanabe, Taniguchi,
  Kroner, and Imamoğlu]{Shimazaki_2020}
Yuya Shimazaki, Ido Schwartz, Kenji Watanabe, Takashi Taniguchi, Martin Kroner,
  and Ataç Imamoğlu.
\newblock Strongly correlated electrons and hybrid excitons in a moiré
  heterostructure.
\newblock \emph{Nature}, 580\penalty0 (7804):\penalty0 472—477, April 2020.
\newblock ISSN 0028-0836.
\newblock \doi{10.1038/s41586-020-2191-2}.
\newblock URL \url{https://doi.org/10.1038/s41586-020-2191-2}.

\bibitem[Tao et~al.(2022)Tao, Zhang, Zhu, He, Yang, Lu, and Wei]{Tao_2022}
Shengdan Tao, Xuanlin Zhang, Jiaojiao Zhu, Pimo He, Shengyuan Yang, Yunhao Lu,
  and Su-Huai Wei.
\newblock Designing ultra-flat bands in twisted bilayer materials at large
  twist angles: Theory and application to two-dimensional indium selenide.
\newblock \emph{Journal of the American Chemical Society}, 144, 02 2022.
\newblock \doi{10.1021/jacs.1c11953}.

\bibitem[Tarnopolsky et~al.(2019)Tarnopolsky, Kruchkov, and
  Vishwanath]{Origin_Magic_2019}
Grigory Tarnopolsky, Alex~Jura Kruchkov, and Ashvin Vishwanath.
\newblock Origin of magic angles in twisted bilayer graphene.
\newblock \emph{Phys. Rev. Lett.}, 122:\penalty0 106405, Mar 2019.
\newblock \doi{10.1103/PhysRevLett.122.106405}.
\newblock URL \url{https://link.aps.org/doi/10.1103/PhysRevLett.122.106405}.

\bibitem[Tokura et~al.(2017)Tokura, Kawasaki, and Nagaosa]{tokura2017emergent}
Yoshinori Tokura, Masashi Kawasaki, and Naoto Nagaosa.
\newblock Emergent functions of quantum materials.
\newblock \emph{Nature Physics}, 13\penalty0 (11):\penalty0 1056--1068, 2017.

\bibitem[Uri et~al.(2020)Uri, Grover, Cao, Crosse, Bagani, Rodan-Legrain,
  Myasoedov, Watanabe, Taniguchi, Moon, et~al.]{uri2020}
Aviram Uri, Sameer Grover, Yuan Cao, John~A Crosse, Kousik Bagani, Daniel
  Rodan-Legrain, Yuri Myasoedov, Kenji Watanabe, Takashi Taniguchi, Pilkyung
  Moon, et~al.
\newblock Mapping the twist-angle disorder and landau levels in magic-angle
  graphene.
\newblock \emph{Nature}, 581\penalty0 (7806):\penalty0 47--52, 2020.

\bibitem[Wong et~al.(2020)Wong, Nuckolls, Oh, Lian, Xie, Jeon, Watanabe,
  Taniguchi, Bernevig, Yazdani, and et~al.]{wong2020}
Dillon Wong, Kevin~P. Nuckolls, Myungchul Oh, Biao Lian, Yonglong Xie, Sangjun
  Jeon, Kenji Watanabe, Takashi Taniguchi, B.~Andrei Bernevig, Ali Yazdani, and
  et~al.
\newblock Cascade of electronic transitions in magic-angle twisted bilayer
  graphene.
\newblock \emph{Nature}, 582\penalty0 (7811):\penalty0 198–202, 2020.
\newblock \doi{10.1038/s41586-020-2339-0}.

\bibitem[Zhang and Tadmor(2017)]{ZHANG_Tadmor_2017}
Kuan Zhang and Ellad~B. Tadmor.
\newblock Energy and moiré patterns in 2d bilayers in translation and
  rotation: A study using an efficient discrete–continuum interlayer
  potential.
\newblock \emph{Extreme Mechanics Letters}, 14:\penalty0 16--22, 2017.
\newblock ISSN 2352-4316.
\newblock \doi{https://doi.org/10.1016/j.eml.2016.10.010}.
\newblock URL
  \url{https://www.sciencedirect.com/science/article/pii/S2352431616301821}.
\newblock Mechanics and Mechanical Behavior of 2D Materials – Graphene and
  Beyond.

\bibitem[Zhang and Tadmor(2018)]{ZHANG_Tadmor_2018}
Kuan Zhang and Ellad~B. Tadmor.
\newblock Structural and electron diffraction scaling of twisted graphene
  bilayers.
\newblock \emph{Journal of the Mechanics and Physics of Solids}, 112:\penalty0
  225--238, 2018.
\newblock ISSN 0022-5096.
\newblock \doi{https://doi.org/10.1016/j.jmps.2017.12.005}.
\newblock URL
  \url{https://www.sciencedirect.com/science/article/pii/S0022509617310153}.

\bibitem[Zhao et~al.(2020)Zhao, Yang, Zhang, and Wei]{Zhao_2020}
Xing-Ju Zhao, Yang Yang, Dong-Bo Zhang, and Su-Huai Wei.
\newblock Formation of bloch flat bands in polar twisted bilayers without magic
  angles.
\newblock \emph{Phys. Rev. Lett.}, 124:\penalty0 086401, Feb 2020.
\newblock \doi{10.1103/PhysRevLett.124.086401}.
\newblock URL \url{https://link.aps.org/doi/10.1103/PhysRevLett.124.086401}.

\bibitem[Zhao et~al.(2021)Zhao, Yang, Zhang, and Wei]{Zhao_2021}
Xing-Ju Zhao, Yang Yang, Dong-Bo Zhang, and Su-Huai Wei.
\newblock Flat bands in twisted bilayers of polar two-dimensional
  semiconductors.
\newblock \emph{Phys. Rev. Mater.}, 5:\penalty0 014007, Jan 2021.
\newblock \doi{10.1103/PhysRevMaterials.5.014007}.
\newblock URL \url{https://link.aps.org/doi/10.1103/PhysRevMaterials.5.014007}.

\bibitem[Zhou et~al.(2015)Zhou, Han, Dai, Sun, and
  Srolovitz]{zhouVanWaalsBilayer2015}
Songsong Zhou, Jian Han, Shuyang Dai, Jianwei Sun, and David~J. Srolovitz.
\newblock Van der {{Waals}} bilayer energetics: {{Generalized}} stacking-fault
  energy of graphene, boron nitride, and graphene/boron nitride bilayers.
\newblock \emph{Phys. Rev. B}, 92\penalty0 (15):\penalty0 155438, October 2015.
\newblock ISSN 1098-0121, 1550-235X.
\newblock \doi{10.1103/PhysRevB.92.155438}.

\end{thebibliography}
\end{document}